\documentclass[aps,prc,showpacs,showkeys,twoside,twocolumn,byrevtex,floatfix]{revtex4-2}
\usepackage{graphicx} %
\usepackage{color}    %

\usepackage{amsmath,amssymb,amsfonts}
\usepackage{bm}
\usepackage{mathtools}

\usepackage{comment}

\usepackage{hyperref}
\hypersetup{
  bookmarksnumbered,%
  colorlinks,%
  linkcolor=blue,%
  citecolor=blue,%
  urlcolor=blue,%
  plainpages=false,%
  pdfstartview=FitH
}

\newcommand{\nc}{\newcommand}           
\nc{\vc}[1]     {\mbox{\boldmath $#1$}} 
\nc{\bras}[1]   {\langle#1|}            
\nc{\kets}[1]   {|#1\rangle}            
\nc{\bra}       {\langle}            
\nc{\ket}       {\rangle}            
\nc{\hO}        {\hat{O}}           
\nc{\HO}        {\hat{O}}   
\nc{\wtil}      {\widetilde}            
\nc{\EV}[1]     {\langle #1 \rangle} 
\nc{\red}[1]    {\textcolor{black}{#1}}  

\begin{document}

\title{Physics of many-body resonances with complex scaling and applications to light unstable nuclei}

\author{Takayuki Myo}
\email{takayuki.myo@oit.ac.jp}
\affiliation{General Education, Faculty of Engineering, Osaka Institute of Technology, Osaka, Osaka 535-8585, Japan}
\affiliation{Research Center for Nuclear Physics (RCNP), Osaka University, Ibaraki, Osaka 567-0047, Japan}

\date{\today}

\begin{abstract} 
  We review the exotic phenomena in light unstable nuclei with a focus on many-body resonances,
  which can decay into more than two constituents, and are frequently observed in unstable nuclei above the three-body threshold energy.
  The complex scaling transformation of the Schr\"odinger equation is a powerful method for describing many-body resonances,
  because it separates the continuum spectra into resonant and non-resonant continuum ones.
  Since the asymptotic wave functions of the resonances are regularized in the complex scaling,
  many-body resonances are described using the $L^2$ basis functions in the eigenvalue problem. 
  The properties of many-body resonances can then be discussed in the same way as those of the bound states.
  We apply the complex scaling to the system consisting of a stable nucleus and valence nucleons
  and investigate many-body resonances in neutron-rich and proton-rich light nuclei.
  Using the eigenstates obtained with the complex scaling, we construct the extended completeness relation and the Green's function.
  They are used to calculate the level densities and the general transition strengths into many-body unbound states.
  We also discuss the interpretation of the complex expectation values associated with resonances, which remains an open problem.
  We propose a possible scheme for it in terms of the complex-scaled Green's function. 
\end{abstract}


\maketitle


\section{Introduction}\label{sec:intro}

Quantum resonances are a general phenomenon in many fields of physics,
including atomic and molecular physics, nuclear and hadron physics, and condensed matter physics \cite{ho83,moiseyev98,hatano08,moiseyev11,hosaka16}.
The physics of resonances is closely related to non-Hermitian physics,
because resonances are often described as eigenstates with complex energies \cite{PTEP2020}.
A complex energy state was firstly introduced by Gamow \cite{gamow28,gamow29}
to explain the $\alpha$ decay phenomenon of nuclei, which exhibits a time dependence of the probability in the exponential form, $e^{-\lambda t}$,
using a decay constant, $\lambda=\Gamma/\hbar$ with a decay width $\Gamma$.
The real and imaginary parts of the complex energy, $E_r-i\Gamma/2$, describe the energy and decay probability of
the emitted $\alpha$ particle, respectively.
The decay constant $\lambda$ can be explained by the tunneling effect of the $\alpha$ particle
with the energy $E_r$ through a Coulomb potential barrier between the $\alpha$ particle and the daughter nucleus.
This description of the decay process is based on a time-dependent picture of the wave function.
On the other hand, Siegert described a decaying state as the one expressed by the outgoing wave asymptotically \cite{siegert39}.
The description by Siegert is based on a stationary picture. The complex-energy state can then be obtained as a solution to
the time-independent Schr\"odinger equation under the boundary condition of the outgoing wave.

In the quantum scattering theory, resonances have been investigated
by calculating the complex eigenenergies as poles of the scattering matrix ($S$-matrix) \cite{humblet61,lane58}.
However, it is generally difficult to treat many-body resonances and non-resonant continuum states,
because a many-body system can be divided into its various constituents in many different ways.
These divisions produce various complex thresholds, which give rise to many-body continuum states and many-body resonances.
Here, we refer to resonances that decay into more than two constituents as ``many-body resonances''.
The properties of many-body resonances are essential to understanding the unbound states of the finite many-body systems such as atomic nuclei.

\begin{figure}[b]
\centering
\includegraphics[width=8.5cm,bb=0 0 439 332]{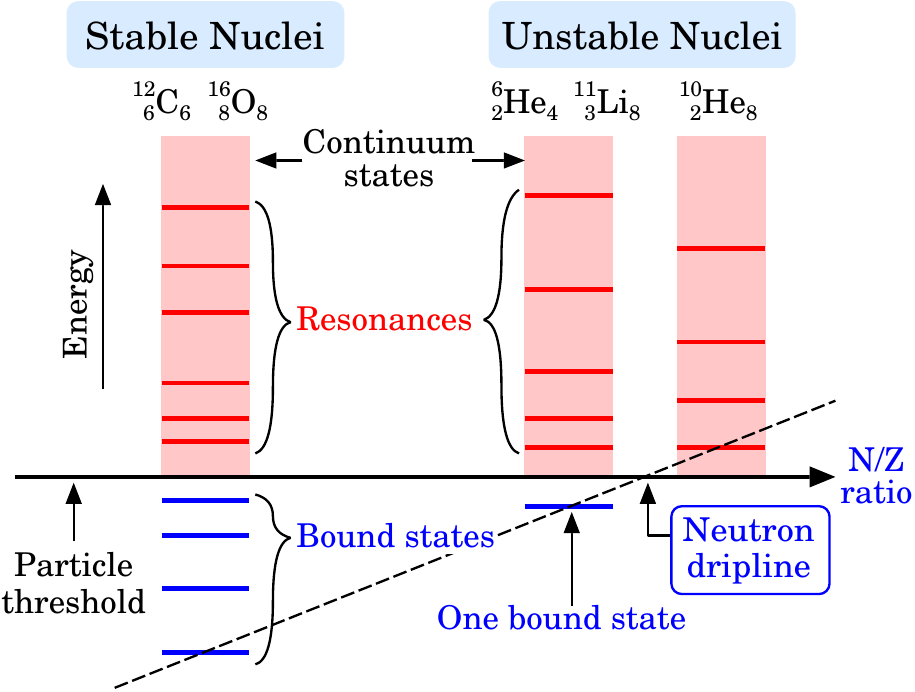}
\caption{
  Schematic diagram to illustrate the relation between the ground states and
  the thresholds in the energy levels from stable (left, $Z \approx N$) to neutron-rich (right, $Z \ll N$) nuclei.
  The notations of nuclei is $^A_Z {\rm X}_N$ with the mass number $A=Z+N$.
  Diagonal dashed line connects the ground states of nuclei.
}
  \label{fig:unbound}
\end{figure}

The properties of the unbound states are fundamental to the nuclear structure and reactions.
In this review, we focus on the light unstable nuclei with the regions that are either neutron-rich ($Z \ll N$) or proton-rich ($Z \gg N$),
which are radioactive.
In the field of unstable nuclear physics, starting from the discovery of the neutron halo structure, which shows a very wide distribution
of a few valence neutrons as seen in $^6_2$He$_4$ and $^{11}_{~3}$Li$_8$ \cite{tanihata85}, 
experimental developments have revealed various interesting phenomena related to the unbound states of nuclei \cite{tanihata96,tanihata13,nakamura17}.
In unstable nuclei, a few extra nucleons are weakly bound to the system with small binding energies of around 1 MeV.
This is in contrast to the average value of 8 MeV per nucleon for stable nuclei.
This indicates that the nucleons can easily be released with low excitation energies,
and that the lowest threshold position is very close to the ground state, as shown in Fig. \ref{fig:unbound}.
For example, $^6$He and $^{11}$Li only have one bound state as their ground states,
while the ground state of $^{10}_{~2}$He$_8$ is a resonance that decays into the $^8_2$He$_6$+$n$+$n$ three-body system.
An interesting feature of unstable nuclei is their Borromean nature of a three-body system,
in which no two-body subsystems have bound states, yet the three-body system is bound.  
The Borromean nuclei such as $^6$He=$\alpha$+$n$+$n$ and $^{11}$Li=$^9_3$Li$_6$+$n$+$n$ have a few bound states and
their lowest thresholds are three-body, not two-body.
Therefore most of the excited states in unstable nuclei are unbound and observed as ``many-body resonances'' \cite{aoyama06,ikeda10,horiuchi12}.
There are many Borromean nuclei observed in the light neutron-rich region; $^8_2$He$_6$, $^{14}_{~4}$Be$_{10}$, $^{17}_{~5}$B$_{12}$, $^{19}_{~5}$B$_{14}$,
and $^{22}_{~6}$C$_{16}$ \cite{nndc}.

\begin{figure}[b]
\centering
\includegraphics[width=7.0cm,bb=0 0 342 327]{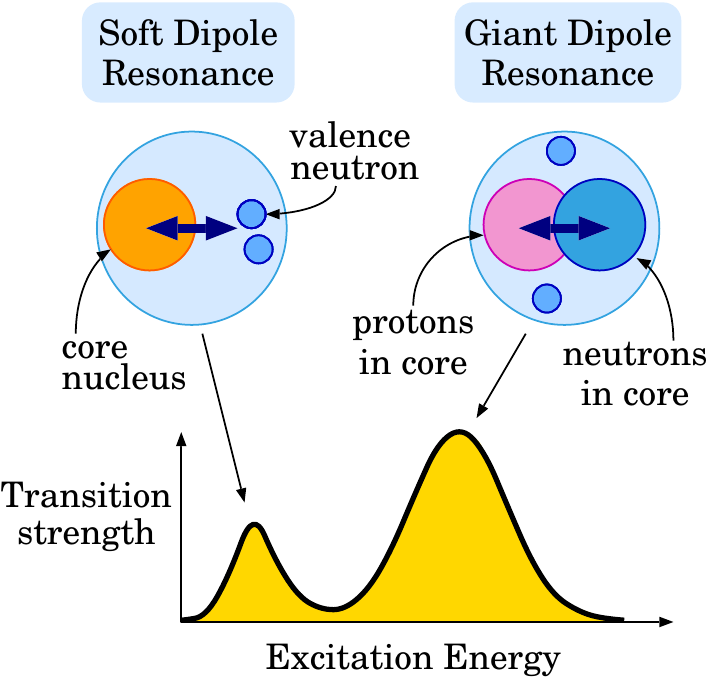}
\caption{A schematic spectrum of the excitations of the soft dipole resonance and giant dipole resonance in two-neutron halo nuclei with core+$n$+$n$.
  Shown above in the left is a soft dipole resonance (SDR) \cite{hansen87,ikeda88,ikeda92}
  and the one in the right is the giant dipole resonance (GDR).}
\label{fig:soft}
\end{figure}

If there is a neutron halo structure, one can expect an exotic excitation mode originating from the excess neutrons, such as for $^{6}$He and $^{11}$Li.
If the $\alpha$ particle or $^9$Li nucleus (core nuclei) is surrounded by a wide neutron halo region,
the core nucleus may oscillate dipolarly in the neutron sea, as shown in Fig. \ref{fig:soft}.
This is called a soft dipole resonance (SDR), which might be excited by the photo-disintegration
with a strong dipole transition \cite{hansen87,ikeda88,ikeda92}.
The excitation energy of SDR is expected to be around a few MeV as compared with the standard giant dipole resonance (GDR),
in which the protons and neutrons oscillate in opposite directions and the excitation energy is a few tens of MeV.
The soft dipole resonance is one of the characteristic excitations of unstable nuclei induced by the excess neutrons \cite{adrich05,paar07}.
Many experiments have been conducted to investigate the exotic dipole excitations of unstable nuclei \cite{aumann99,nakamura06,kanungo15},
some of which show a low-lying enhancement in the transition strength just above the lowest threshold.
However, it is still being discussed whether the observed enhancement is caused by the soft dipole resonance.

Therefore, a new approach to many-body resonances is desirable for developing the unstable nuclear physics,
and the complex scaling is a promising approach for this purpose
\cite{nuttall69,aguilar71,balslev71,connor83,ho83,moiseyev98,moiseyev11,carbonell14,aoyama06,myo14a,myo20}.
It is known that the resonance wave function with a complex eigenenergy diverges exponentially at large distances
according to the Siegert condition \cite{siegert39},
but, this singularity disappears in the complex scaling as is explained in the Sect. \ref{sec:method}.
In this method, the resonance wave function is described by the square-integrable $L^2$ functions, as are bound states.
Using the appropriate basis functions, one can uniquely obtain many-body resonances and also the non-resonant continuum states,
both of which are necessary to understand the scattering phenomena of unstable nuclei.

The aim of this paper is to provide a brief review of many-body resonances in light unstable nuclei and the complex scaling.
We demonstrate several applications to many-body resonant and continuum states of light unstable nuclei.
The complex scaling has been brought by Hartree et al. (see Ref. \cite{connor83}), Nuttall and Cohen \cite{nuttall69}, and Aguilar, Combes, and Balslev \cite{aguilar71,balslev71}.
The boundary condition of the outgoing wave is automatically imposed for the resonances.
The resonance energy $E_r$ and decay width $\Gamma$ are obtained directly from the complex energy eigenvalues of $E_r-i\Gamma/2$.

Resonances are often described with the scattering solutions of the Schr\"odinger equation.
The properties of these states have been discussed in the nuclear reaction theory \cite{humblet61}.
In a different approach, Berggren has considered to describe the resonances as the extension of the bound state concept \cite{berggren68,berggren93}.
He introduced the extended completeness relation,
in which the resonance poles are explicitly included using the deformed contours on the complex momentum plane.
This method of complex contour deformation are utilized to prepare the single-particle basis states to describe weakly bound and resonant states
in nuclei \cite{betan02,michel08,okolowicz03,michel22}.

Among the unstable nuclei, neutron-rich nuclei have been extensively studied \cite{tanihata13,nakamura17}.
In neutron-rich nuclei, multi-neutron resonances above the core nucleus should be described to clarify their exotic properties.
In proton-rich nuclei, the Coulomb interaction produces a repulsive effect in the energies, and
the number of resonances is generally larger than in the neutron-rich nuclei \cite{charity11,sobotka24,koyama24}.
The complex scaling can be used to study many-body resonance phenomena in both neutron-rich and proton-rich nuclei. 
In this article, we show some examples; the neutron-rich He isotopes of $^{5\mbox{-}8}_{~~2}$He and their mirror nuclei,
the proton-rich $^5_3$Li$_2$, $^6_4$Be$_2$, $^7_5$B$_2$, and $^8_6$C$_2$, which are the $N=2$ isotone and all unbound \cite{myo21,myo22a,myo22b,myo23a}.

Complex scaling provides with a natural extension of the ordinary completeness relation,
which consists of bound and scattering (unbound) states, to include bound, resonance, and non-resonant continuum states.
Using the extended completeness relation for many-body scattering solutions,
we can construct the complex-scaled Green's function of the system.
The Green's function is essential for obtaining the strength functions excited from the ground state to the many-body scattering states
\cite{myo98,myo01,myo07b}. 
We apply the complex scaling to extend the completeness relation,
and introduce the complex-scaled Green's function to calculate the transition strengths into the many-body resonances.

Similar to the eigenenergies, it is known that the expectation values of a Hermitian operator for resonances can be complex numbers.
The physical interpretation of these complex values is a long-standing problem
\cite{berggren96,burgers96,homma97,sekihara13,dote18,myo14b,michel22,myo23b}.
Recently, we proposed a possible scheme for it ;
we utilize the Green's function of resonances in the strength function and
the complex expectation values become the source of the strength function on the real energy axis. 
We applied this scheme to the radii of resonances, the interpretation of which has been discussed
in various fields of physics \cite{burgers96,homma97,sekihara13,dote18,myo14b}.

The last section presents a summary.
The remaining subjects of many-body resonances and the complex scaling are presented as future perspectives
in relation to the unified description of quantum many-body unbound states.

\section{Unified description of bound, resonant and continuum states} \label{sec:method}

\subsection{Complex scaling} \label{sec:ABC}

\begin{figure}[b]
\centering
\includegraphics[width=8.6cm,bb=0 0 505 213]{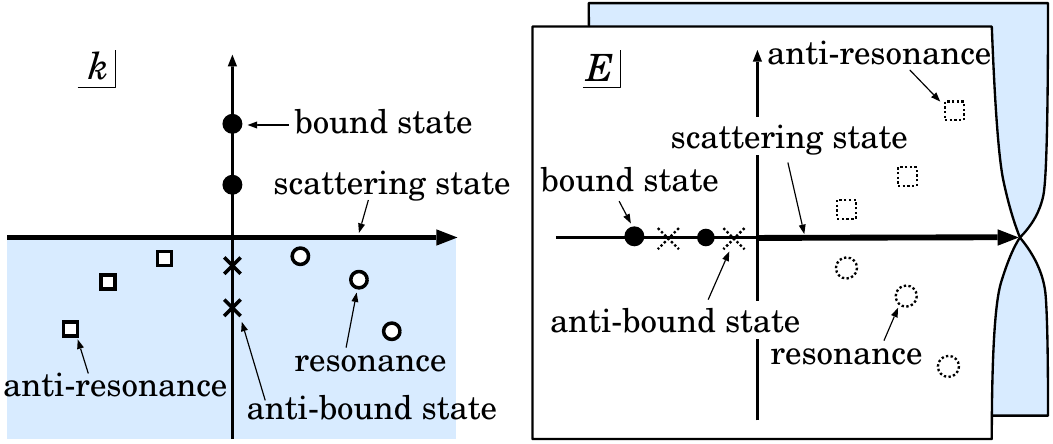}
\caption{$S$-matrix poles in the momentum (left) and energy (right) planes. 
The thick horizontal arrows indicate the scattering states, which have a real momentum and a positive energy.}
\label{fig:sec2_pole-k}
\end{figure}

This section explains the fundamental framework of the complex scaling from the two-body system to many-body systems.
Figure~\ref{fig:sec2_pole-k} shows the schematic pole distribution of the $S$-matrix for a two-body system with a single channel
on the complex momentum ($k$) and energy ($E$) planes.
The energy plane consists of two Riemann sheets. 
In the complex momentum plane, the discrete solutions are classified as bound states, anti-bound states, resonances, and anti-resonances
with the corresponding momenta:
\begin{center}
\renewcommand{\arraystretch}{1.50}
\begin{tabular}{c|lcc}
     bound states       &~~$k_{\rm B} = i\gamma_b$ \\ \hline
     anti-bound states  &  $k_{\rm AB}=-i\gamma_{ab}$  \\  \hline
     resonances         &~~$k_{\rm R} =  \kappa_r-i\gamma_r$   \\  \hline
     anti-resonances    &  $k_{\rm AR}= -\kappa_r-i\gamma_r$ \\ 
\end{tabular}\\ 
\end{center}
where $\kappa_r$, $\gamma_r$, $\gamma_b$, and $\gamma_{ab}$ are all positive numbers.
We assume that the asymptotic wave functions are proportional to $e^{ik_{\rm p}r}$, 
where the momentum $k_{\rm p}=k_{\rm B}$,~$k_{\rm AB}$, $k_{\rm R}$, and $k_{\rm AR}$, respectively. 
Therefore, only the bound states have a damping form of the radial wave function, which asymptotically approaches 
$\psi_{k_{\rm B}} \sim e^{-\gamma_b r}$.
Additionally, there exists the virtual states on the negative imaginary axis for the $s$-wave case.

The resonance wave function, $\Phi_{\rm R}$, has a complex eigenenergy, $E_{\rm R}=E_r-i\Gamma/2=\hbar^2k_{\rm R}^2/(2\mu)$,
with a reduced mass $\mu$.
The adjoint state of the resonance is the anti-resonance $\Phi_{\rm AR}$
with the eigenenergy of $E_{\rm AR}=E_r+i\Gamma/2=E_{\rm R}^*$ and the momentum $k_{\rm AR}=-k_{\rm R}^*$.
This state is also called a capturing state or a growing state because of the time dependence of the probability
as $e^{\Gamma t/\hbar}$ \cite{berggren82,hatano08}.
The resonance and anti-resonance form a bi-orthogonal relation \cite{berggren68}, and their radial components have a relation of 
$\Phi_{\rm AR,radial}=\Phi_{\rm R,radial}^*$. One often uses the notation of $\Phi_{\rm AR}$ as $\wtil \Phi_{\rm R}$ \cite{berggren68}.
This relation is used in calculating of the matrix elements of resonances.
For the continuum state $\Phi_k$ with a complex momentum $k$, the adjoint state $\wtil \Phi_k$ has the momentum $k^*$ \cite{berggren68,myo97}.

We will explain the complex scaling proposed mathematically by Aguilar, Balslev, and Combes \cite{aguilar71,balslev71}.
They introduced the transformation operator $U(\theta)$, which has a real scaling angle $\theta$, for the spatial coordinates $\{\vc{r}_i\}$
and momenta $\{\vc{k}_i\}$ of all particles, where the particle index $i=1,\ldots, n$, in many-body systems as,
\begin{eqnarray}
  U(\theta)\vc{r}_iU^{-1}(\theta)&=&\vc{r}_i \, e^{ i\theta},
  \\
  U(\theta)\vc{k}_iU^{-1}(\theta)&=&\vc{k}_i \, e^{-i\theta},
\end{eqnarray}
where $U(\theta)\,U^{-1}(\theta)=1$.
We usually employ the same $\theta$ for all particles.
The Schr\"odinger equation for the many-body wave function $\Phi$ is given as follows:
\begin{eqnarray}
  H\,\Phi(\vc{r}_1,\ldots,\vc{r}_n)=E\, \Phi(\vc{r}_1,\ldots,\vc{r}_n),
\end{eqnarray}
with the Hamiltonian $H$ having the interaction term $V$ and subtracting the center-of-mass kinetic energy $T_{\rm c.m.}$, defined as: 
\begin{eqnarray}
H=\sum_{i=1}^nT_i-T_{\rm c.m.}+\sum_{i<j}^nV(\vc{r}_i-\vc{r}_j),
\end{eqnarray}
which is transformed as 
\begin{eqnarray}
  H^\theta \Phi^\theta=E^\theta \Phi^\theta,
  \label{eq:CSEQ}
\end{eqnarray}
where the complex-scaled Hamiltonian $H^\theta$ is given as 
\begin{eqnarray}
  H^\theta&=&U(\theta)HU^{-1}(\theta)
  \nonumber\\
  &=&
  e^{-2i\theta}\left\{\sum_{i=1}^nT_i-T_{\rm c.m.}\right\}
  +\sum_{i<j}^n V\left((\vc{r}_i-\vc{r}_j)e^{i\theta}\right),
  \nonumber\\
  \Phi^\theta&=&U(\theta)\Phi(\vc{r}_1,\ldots,\vc{r}_n)
  =e^{if\theta/2} \Phi(\vc{r}_1e^{i\theta},\ldots,\vc{r}_ne^{i\theta}).
  \nonumber\\
  \label{eq:CSWF}
\end{eqnarray}
The many-body wave function $\Phi^\theta$ is the internal wave function excluding the center-of-mass effect,
and $f=3n-3$ is the number of internal degrees of freedom.
The factor $e^{if\theta/2}$ comes from the Jacobian of the coordinate $\vc{r}_i$ in the matrix elements.

\begin{figure}[b]
\centering
\includegraphics[width=8.5cm,bb=0 0 459 221]{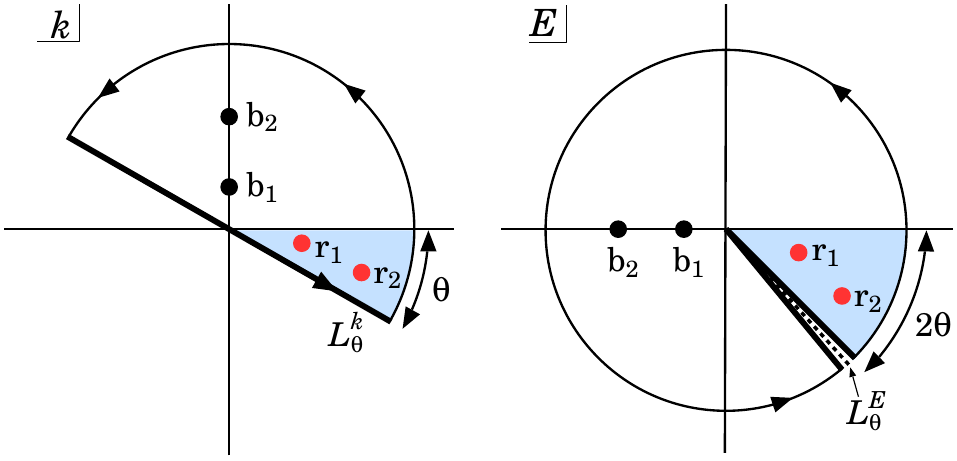}
\caption{
  Cauchy integral contours in the momentum and energy planes for the completeness relation with complex scaling ($\theta>0$).
  The solid circles $b_1$, $b_2$ and $r_1$, $r_2$ are the poles of bound and resonant states, respectively.
  The continuum states are obtained along the rotated axis shown by $L^k_\theta$ and $L^E_\theta$ with thick lines.
}
\label{fig:pole}
\end{figure}

The asymptotic wave functions for the poles with momenta $k_{\rm p}$ are expressed using the outgoing waves
in the radial part as $e^{ik_{\rm p} r e^{i\theta}}$ in the two-body system. 
The resonance wave functions, which originally exhibit divergent behavior due to the presence of $\gamma_r$ as
$e^{ik_{\rm R}\cdot r}=e^{i(\kappa_r-i\gamma_r)r}=e^{i\kappa_r r}\cdot e^{\gamma_r r}$, are transformed in the complex scaling as
\begin{eqnarray}
    e^{ik_{\rm R}\cdot re^{i\theta}}
&=& e^{ir(\kappa_r-i\gamma_r) (\cos \theta +i\sin \theta )} 
    \nonumber \\
&=& e^{(-\kappa_r\sin{\theta}+\gamma_r\cos{\theta})r}\cdot e^{i(\kappa_r\cos{\theta}+\gamma_r\sin{\theta})r}. 
\end{eqnarray}
From this equation, it is shown that the divergent behavior of the resonance wave function is regularized when 
the scaling angle $\theta$ is greater than the specific angle $\theta_r=\tan^{-1}(\gamma_r/\kappa_r)=|\arg(k_{\rm R})|$
at the resonance position $k_{\rm R}=\kappa_r-i\gamma_r$. 
One can obtain the resonance wave function with $\theta>\theta_r$, in the damping form together with the bound states,
as shown in Fig. \ref{fig:pole}.
For anti-resonances with a complex momentum $k_{\rm AR}=-\kappa_r-i\gamma_r$, 
the sign of $\kappa_r$ changes, so one takes the negative value of $\theta<-\theta_r$ to put the state in the damping form.

The properties of the solutions of the complex-scaled Schr\"odinger equation are summarized in the so-called ABC theorem \cite{aguilar71,balslev71} as:
\begin{enumerate}
\itemsep=0.1cm
\item The energies of the bound states remain unchanged under scaling transformations.
\item The resonance wave functions are described by the square-integrable $L^2$ functions, similar to bound states. 
\item The continuum spectra begin at the threshold energies of the system separated into the subsystems.
  The spectra are rotated clockwise by the angle of $2\theta$ from the positive real energy axis in the complex energy plane.
\end{enumerate}
When we solve Eq.~(\ref{eq:CSEQ}), we obtain the complex energy eigenvalues that are categorized 
into bound, resonant, and non-resonant continuum states. 
A schematic distribution of the eigenvalues of $H^\theta$ in the ABC theorem is shown in Fig.~\ref{fig:pole}.

We assume that $\Phi_{\rm R}$ is an eigensolution of resonance in Eq.~(\ref{eq:CSWF}) with a momentum $k_{\rm R}$.
Its conjugate solution is expressed by $\wtil{\Phi}_{\rm R}$ for the bi-orthogonal state \cite{berggren68,aoyama06}. 
Unlike the ordinary Hermite product,
in the bi-orthogonal relation, the matrix elements for the arbitrary operator in the complex scaling are expressed as 
\begin{eqnarray}
	\bra \wtil{\Phi}_{\rm R}|\hat{O} | \Phi_{\rm R}\ket
&=&	\bra U(\theta)\wtil{\Phi}_{\rm R}| U(\theta)\hat{O}U^{-1}(\theta) | U(\theta)\Phi_{\rm R}\ket
        \nonumber\\
&=&     \bra \wtil{\Phi}^\theta_{\rm R}|\hat{O}^\theta | \Phi^\theta_{\rm R}\ket ,
        \label{eq:CSME}
        \nonumber \\
        \hat{O}^\theta 
&=&     U(\theta)\hat{O} U^{-1}(\theta), 
\end{eqnarray}
where $\hat{O}^\theta$ is the complex-scaled operator. 
To express the resonance wave function in Eq.~(\ref{eq:CSWF}),
we often expand the wave functions $\Phi^\theta_{\rm R}(\vc{r})$ in terms of a finite number of
the $L^2$ basis functions, $\{u_i(\vc{r})\}$, with the basis index $i=1,2,\ldots, N$,
such as harmonic oscillator basis functions and the Gaussian basis functions \cite{aoyama06}.
Here, we briefly explain the method for the single-channel case of a two-particle system with a relative coordinate $\vc{r}$.
The radial part of $u_i(\vc{r})$ should be real and the basis number $N$ is determined so that the solutions converge. 
This expansion is straightforwardly applicable to many-body systems. 
The complex-scaled wave function and the complex-scaled energy $E^\theta$ are given as follows:
\begin{eqnarray}
      \Phi^\theta_{\rm R}(\vc{r})
&=&   \sum_i^{N} c_{i}(\theta)\, u_i(\vc{r}),
      \label{eq:base}
      \\
      \wtil{\Phi}^\theta_{\rm R}(\vc{r})
&=&   \sum_i^{N} c^*_{i}(\theta)\, \wtil{u}_i(\vc{r}),
      \\
      E^\theta &=& \dfrac{\bra \wtil{\Phi}^\theta_{\rm R}|H^\theta | \Phi^\theta_{\rm R}\ket}{\bra \wtil{\Phi}^\theta_{\rm R}|\Phi^\theta_{\rm R}\ket}.
\end{eqnarray}
The bi-variational principle for the complex-scaled energy $\delta E^\theta~=~0$ leads to the generalized eigenvalue problem \cite{moiseyev98}.
\begin{eqnarray}
      \sum_i^{N}( H_{ij}^\theta\ - N_{ij}) c_j(\theta) 
~=~   E^\theta c_i(\theta),
      \label{eq:eigen}
      \\
    H_{ij}^\theta
~=~ \bra \wtil{u}_i | H^\theta | u_j \ket ,\quad
    N_{ij}
~=~ \bra \wtil{u}_i | u_j \ket ,
\end{eqnarray}
where $H_{ij}^\theta$ are the matrix elements of the complex-scaled Hamiltonian.
These equations provide the stationary condition for the eigenenergies of the resonances with respect to the scaling angle $\theta$.
By solving the eigenvalue problem, one obtains the coefficients $\{c_i(\theta)\}$ and the discrete energy spectra $E^\theta$.
In the basis expansion form of Eq.~(\ref{eq:base}),
the non-resonant continuum states are also obtained in the discretized spectra
with the eigenvalues on the $2\theta$ line on the complex energy plane.

After that, we calculate the various physical quantities associated with resonances.
The matrix elements for the arbitrary operator in Eq.~(\ref{eq:CSME}) are calculated using the coefficients $c_i(\theta)$ as
\begin{eqnarray}
  \bra \wtil{\Phi}^\theta_{\rm R}|\hat{O}^\theta | \Phi^\theta_{\rm R}\ket 
  =     \sum_{i,j}^N c_i(\theta)\, c_j(\theta)\,  \bra \wtil{u}_i | \hat{O}^\theta | u_j \ket, 
  \label{eq:csme}
\end{eqnarray}
where the complex conjugate is not taken for the coefficient $c_i(\theta)$, namely the radial part of $\widetilde{\Phi}^\theta_{\rm R}$,
due to the biorthogonal property \cite{berggren68,myo97,moiseyev98}.

In the complex scaling, the transformed interaction $ V(\vc{r}e^{i\theta}) = U(\theta) V(\vc{r}) U^{-1}(\theta) $  must maintain analyticity,
which limits the form of the interaction.
For the Gaussian interaction, $e^{-ar^2}$, the complex scaling is applicable for $\displaystyle |\theta| < \pi/4$
because
\begin{eqnarray}
  U(\theta)\,e^{-ar^2}U^{-1}(\theta)
  &=&\exp(-ar^2 e^{2i\theta})
  \nonumber\\
  &=&\exp(-ar^2 \cos 2\theta-iar^2\sin 2\theta),
  \nonumber\\
\end{eqnarray}
and it satisfies $\cos2\theta>0$ to avoid the divergence at $r\to\infty$.
When the interaction is not analytic, the exterior scaling transformation becomes useful.
This transformation changes the contour of the coordinate integration to avoid the non-analytic region of the interaction.
The details of the method are explained in Refs. \cite{simon79,moiseyev98,moiseyev11}.

\subsection{Three-body resonances} \label{sec:three-Body}

\begin{figure}[b]
\centering
\includegraphics[width=7.0cm,bb=0 0 397 249]{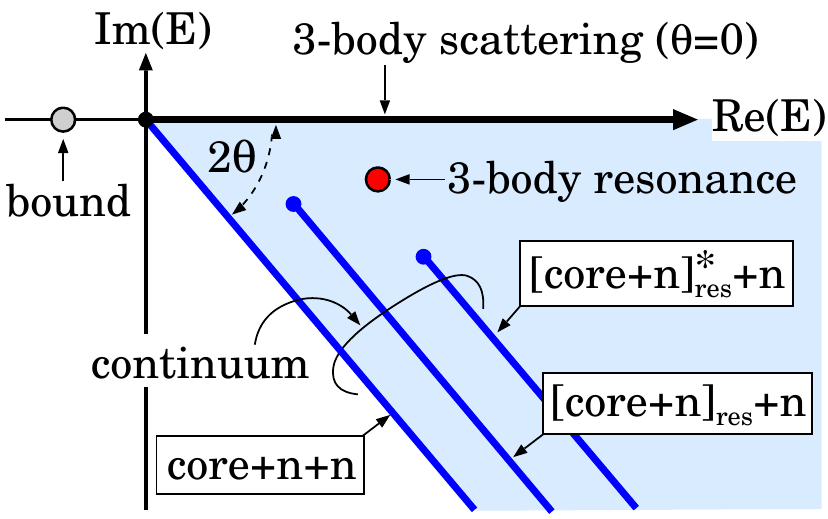}\\[0.4cm]
\includegraphics[width=8.2cm,bb=0 0 362 254]{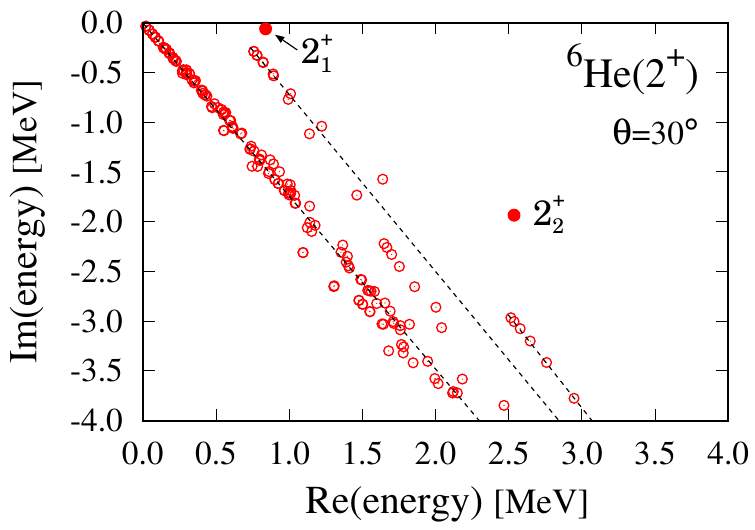}
\caption{
Top: Schematic energy eigenvalue distribution of the Borromean system consisting of [a core nucleus]+$n$+$n$,
where zero energy is the three-body breakup threshold.
The notation of [core+n]$^{(*)}_{\rm res}$ indicates the resonance consisting of a core nucleus and a neutron, 
in which an asterisk means the excited state. 
Bottom: Energy eigenvalues of the $^6$He ($2^+$) states in the complex energy plane,
measured from the $\alpha$+$n$+$n$ threshold with $\theta=30^\circ$ in the complex scaling \cite{myo01,myo20}.
The solid circles indicate the two resonances and
the three schematic dotted lines correspond to the $\alpha$+$n$+$n$ and $^5$He($3/2^-$,$1/2^-$)+$n$ continuum states,
in order from left to right, respectively.}
\label{fig:borromean}
\end{figure}

Complex scaling can be used to describe the many-body resonances and continuum states
and this property greatly extends the applicability of the method.
In this subsection, we discuss the Borromean nuclear three-body system, which is often observed in neutron-rich nuclei
and is described by [a core nucleus]+$n$+$n$ model.
The Borromean nuclear system consists of three constituents of nuclei/nucleons and forms bound states;
however, there are no bound states in any two-body subsystems, i.e. the core+$n$ system can be a resonance,
as shown in the top panel of Fig.~\ref{fig:borromean}. 
This property means that the three-body breakup threshold is the lowest threshold in the system.
All the Borromean nuclei observed so far have very few bound states,
and many excited states are observed as resonances at low excitation energies, as shown schematically in Fig. \ref{fig:unbound}.
Due to the small binding energy of the ground state measured from the lowest threshold, typically around 1 MeV,
Borromean nuclei can easily break apart due to weak perturbation. 
Three-body continuum states appear on the $2\theta$ line starting from the three-body threshold 
in the top panel of Fig.~\ref{fig:borromean}.
The discrete states above the three-body threshold are resonances, including the ground states of the subsystems, 
such as [core+$n$]$_{\rm res}$. 
The two-body continuum spectra of [core+$n$]$_{\rm res}$+$n$ also start from the energies of the two-body resonances, 
which are the thresholds with complex numbers. A last neutron is in a continuum state.
This discussion is extendable to the many-body systems consisting of more than three constituents.

Bottom of Fig. \ref{fig:borromean} shows an example of the eigenvalue distribution of the $^6$He ($2^+$) calculation
using $\alpha$+$n$+$n$ model in the complex scaling \cite{myo01}.
In this case, there is no three-body $2^+$ bound state and the $\alpha$+$n$ system 
forms two $p$-wave resonances of $^5$He, with spin-parity of $J^\pi=3/2^-$ and $1/2^-$.
The schematic three dotted lines indicate the $2\theta$ lines for the rotated two- and three-body continuum states.
In the numerical calculation, the continuum energies are discretized, so they are not completely aligned 
with the $2\theta$ lines \cite{myo01,aoyama06}.
In this distribution, we can identify the locations of the three-body resonances of $^6$He ($2^+_1$, $2^+_2$),
and of two types of continuum states of three-body $^4$He+$n$+$n$ and two-body $^5$He ($3/2^-$, $1/2^-$)+$n$, 
from left to right in the figure.

In the nuclear physics, there have been many applications of
the complex scaling to describe the structures and reactions of nuclear many-body systems
\cite{kruppa88,csoto94,matsumoto10,kikuchi13a,garrido03,yamamoto09,horiuchi13,ohtsubo13,ogata13,zhang22,myo23c}.
Recently, complex scaling has been applied to the resonance calculations of the exotic hadrons \cite{wang22,sakai25}.

\subsection{Extended completeness relation and Green's function}

In ordinary quantum mechanics without complex scaling, bound and scattering (continuum) states construct a complete set,
that is represented by the completeness relation involving the real eigenenergies (momenta) of the Hamiltonian $H$ \cite{newton60} : 
\begin{eqnarray}
\vc{1}&=& \sum_{b=1}^{N_b}|\Phi_b\rangle\langle\Phi_b|+\int_0^\infty dE\,|\Phi_E\rangle\langle\Phi_E|
\nonumber\\
&=&\sum_{b=1}^{N_b}|\Phi_b\rangle\langle\Phi_b|+\int_{-\infty}^\infty dk\,|\Phi_k\rangle\langle\Phi_k|,\label{eq:CR}
\end{eqnarray}
where $\Phi_b$ and $\Phi_E$ ($\Phi_k$) are the bound and continuum states, respectively, and $N_b$ is the number of bound states.
The continuum states $(\Phi_{-k}, \Phi_k)$ in the momentum representation belong to the states on the real $k$ axis.
Integration over the real $k$ axis corresponds to integration along the cut of the first Riemann sheet of the energy plane,
as shown in Fig. \ref{fig:pole} putting $\theta=0$. 
The mathematical proof of the completeness relation in Eq.~(\ref{eq:CR}) was introduced by Newton using Cauchy's theorem \cite{newton60}.

In the complex scaling, the momentum axis rotates down by an angle of $\theta$.
The poles $(r_1,~r_2, \ldots, r_{N^\theta_r})$ of the resonances can enter the semicircle in the Cauchy integration shown in Fig. \ref{fig:pole},
where $N^\theta_r$ represents the number of resonances in the semicircle that has rotated down by $\theta$ in the momentum plane. 
These resonances are explicitly included in the completeness relation of the complex-scaled Hamiltonian $H^\theta$ as follows:
\begin{eqnarray}
  \vc{1}&=&
  \sum_{b=1}^{N_b}            |\Phi^\theta_b\rangle \langle\widetilde{\Phi}^\theta_b|+
  \sum_{r=1}^{N_r^\theta}     |\Phi^\theta_r\rangle \langle\widetilde{\Phi}^\theta_r|+
  \int_{L_\theta^E}dE_\theta\,|\Phi_{E_\theta}\rangle \langle\widetilde{\Phi}_{E_\theta}|
  \nonumber\\
  &=&
  \sum_{b=1}^{N_b}            |\Phi^\theta_b\rangle \langle\widetilde{\Phi}^\theta_b|+
  \sum_{r=1}^{N_r^\theta}     |\Phi^\theta_r\rangle \langle\widetilde{\Phi}^\theta_r|+
  \int_{L_\theta^k}dk_\theta\,|\Phi_{k_\theta}\rangle \langle\widetilde{\Phi}_{k_\theta}|,
  \nonumber\\
  \label{eq:ECR}
\end{eqnarray}
where $\Phi^\theta_b$ and $\Phi^\theta_r$ are the complex-scaled bound and resonant states, respectively.
The tilde in the bra states indicates the biorthogonal states with respect to the ket states satisfying 
$\langle \widetilde{\Phi}^\theta_\nu | \Phi^\theta_{\nu'} \rangle=\delta_{\nu,\nu'}$ with the state indices of $\nu$ and $\nu'$.
The continuum states, $\Phi_{E_\theta}$ and $\Phi_{k_\theta}$, are the solutions obtained on the rotated cut $L_\theta^E$ ($2\theta$ line)
of the Riemann plane and on the rotated momentum axis $L^k_\theta$ ($\theta$ line), respectively.
Hereafter, we refer to the relation in Eq.~(\ref{eq:ECR}) as the extended completeness relation (ECR) \cite{myo98},
which has been proven for single- and coupled-channel systems \cite{giraud03,giraud04}.
There is a recent development of the couple-channel $S$-matrix formulation \cite{yamada22}.

Applying ECR to the calculations of physical quantities enables us to investigate
the contributions from bound, resonant, and non-resonant continuum states separately.
To this end, we present a spectral expansion of the Green's function in the complex scaling, which is given with $H^\theta$ as follows:
\begin{eqnarray}
{\cal G}^\theta(E;\vc{r},\vc{r}')
&=&
U(\theta)\, {\cal G}(E;\vc{r},\vc{r}')\, U^{-1}(\theta)
\nonumber\\
&=&\left\langle\vc{r}\left|\frac{1}{E-H^\theta}\right|\vc{r}'\right\rangle,
\label{eq:green}
\end{eqnarray}
where we use the index $"\theta"$ instead of $(+)$
as the superscript of the outgoing Green's function and we often drop the $i\epsilon$ in the denominator
of the resonance and continuum terms.

By applying the ECR given in Eq.~(\ref{eq:ECR}) to Eq.~(\ref{eq:green}), we obtain the complex-scaled Green's function (CSGF):
\begin{eqnarray}
 {\cal G}^\theta(E;\vc{r},\vc{r}')&=&
  \sum_{b=1}^{N_b}                \frac{|\Phi^\theta_b  \rangle \langle\widetilde{\Phi}^\theta_b  | }{E-E_b+i\epsilon}+
  \sum_{r=1}^{N_r^\theta}         \frac{|\Phi^\theta_r  \rangle \langle\widetilde{\Phi}^\theta_r  | }{E-E_r^{\rm res}+i\Gamma_r/2}  \nonumber\\
&+& \int_{L_\theta^E}dE_\theta \, \frac{|\Phi_{E_\theta}\rangle \langle\widetilde{\Phi}_{E_\theta}| }{E-E_\theta},
\label{eq:CSGF}
\end{eqnarray}
where the resonance eigenenergies are expressed as $E_r^{\rm res}-i\Gamma_r/2$ with the index $r$.
We often represent the non-resonant continuum states using the discretized solutions with a number of $N_c^\theta$ and the index of $c$ as follows:
\begin{eqnarray}
  \int_{L_\theta^E} dE_\theta \,  \frac{|\Phi_{E_\theta} \rangle \langle\widetilde{\Phi}_{E_\theta}|}{E-E_\theta}
  \approx \sum_{c=1}^{N_c^\theta} \frac{|\Phi^\theta_c \rangle \langle\widetilde{\Phi}^\theta_c|}{E-E_c^\theta},
\end{eqnarray}
where the total number of the eigenstates $N=N_b+N_r^\theta+N_c^\theta$ corresponds to the basis number in Eq.~(\ref{eq:base}).
This discretization is useful to describe the many-body continuum states
with the complex eigenenergies of $E_c^\theta$.

It should be noted that the resonance terms are extracted from the original continuum term on the real positive energy axis,
which constructs the resonance structures together with a background.
This continuum term can exhibit rich structures in the coupled-channel and many-body systems.
In these systems, various kinds of continuum states can exist degenerately on the real positive energy axis,
but they are often difficult to distinguish.
In the complex scaling, these states are commonly rotated and distributed along the different $2\theta$ lines
starting from the individual threshold energies, as illustrated in Fig.~\ref{fig:borromean}.
One can distinguish not only the resonances but also the various types of the non-resonant continuum states.
This is an advantage of the complex scaling.

\subsection{Continuum level density}

Continuum level density (CLD), denoted as $\Delta(E)$, is an important quantity in describing quantum scattering phenomena, 
since it is connected with the scattering $S$-matrix  \cite{levine69,tsang75,osborn76,kruppa98}. 
We apply the complex scaling to the CLD and demonstrate its instructive applications to the two-body systems
using the complex-scaled Green's function, which is extendable to many-body cases straightforwardly in the strength function.
We start from the density of states with the real energy $E$, which is defined as follows \cite{levine69}:
\begin{eqnarray}
\rho(E)={\rm Tr}\,\bigl[\delta(E-H)\bigr].
\end{eqnarray}
From the relation 
\begin{eqnarray}
\frac{1}{E^\pm-H}=P\left[\frac{1}{E-H}\right] \mp i\pi\delta(E-H),\label{eq:delta}
\end{eqnarray}
where $E^\pm=E\pm i\epsilon$ with a real and positive $\epsilon$ and the limit of $\epsilon\to 0$ is taken at the final calculation.
We can have 
\begin{eqnarray}
\rho(E)=-\frac{1}{\pi}{\rm Im}\,{\rm Tr}\,\left[\frac{1}{E^+-H}\right].
\end{eqnarray}
The CLD is defined as
\begin{eqnarray}
  \Delta(E)&=&\rho(E)-\rho_0(E)
  \nonumber\\
  &=& -\frac{1}{\pi}{\rm Im}\, \Bigl\{ {\rm Tr}\left[{\cal G}(E^+)-{\cal G}_0(E^+)\right]\Bigr\},
\end{eqnarray}
where the Green's functions are ${\cal G}(E^+)=1/(E^+-H)$ and ${\cal G}_0(E^+)=1/(E^+-H_0)$
for the Hamiltonian $H$ and the asymptotic free Hamiltonian $H_0$, respectively.  

The CLD describes the density of the energy levels, resulting from the interactions in $H$ with a finite range,
and is related to the scattering $S$-matrix, $S(E)$ \cite{levine69}:
\begin{eqnarray}
  \Delta(E)=\frac{1}{2\pi}{\rm Im}\,\frac{d}{dE}\ln\{\det S(E)\}.
\end{eqnarray}
The scattering $S$-matrix for a single-channel system is expressed as $S(E)=e^{2i\delta(E)}$, where $\delta(E)$ is the scattering phase shift.
Therefore, in a single-channel two-body system, we have
\begin{eqnarray}
  \Delta(E)&=&\frac{1}{\pi}\frac{d\delta}{dE},
  \quad
  \delta(E) = \pi\int^{E}_{-\infty}\Delta(E')\,dE'.
\end{eqnarray}
 
Applying the complex scaling to the CLD using the basis expansion method with the basis number $N$, we have
\begin{eqnarray}
  \Delta(E) &=& -\frac{1}{\pi}{\rm Im} \left\{\,
      \sum_{b=1}^{N_b}\frac{1}{E-E_b+i\epsilon}  \right. 
    \nonumber\\
    &+& \sum_{r=1}^{N_r^\theta} \frac{1}{E-E_r^{\rm res}+i\Gamma_r/2  }
    +   \sum_{c=1}^{N_c^\theta} \frac{1}{E-E_c^\theta  }
    \nonumber\\
    &-& \left. \sum_{k=1}^N     \frac{1}{E-E_{0,k}^\theta  }\, \right\},
\end{eqnarray}
where $E_{0,k}^\theta$ with the state index $k$ are the energy eigenvalues of the free Hamiltonian with complex scaling, $H^\theta_0$,
obtained using the same basis functions, $\{u_i\}$, as used for $H^\theta$ in Eq.~(\ref{eq:eigen}).
In principle, $\Delta(E)$ is related to the physical quantity and is therefore independent of $\theta$, but 
in numerical calculations, it may have a slight $\theta$ dependence due to the finite basis number $N$ to represent the eigenstates.
We adopt a sufficiently large $N$ to maintain numerical accuracy and minimize the dependence on $\theta$ in the solutions.
This formulation is general and applicable not only to the nuclear scatterings, but also to the collision of ions \cite{sano24}.

\subsection{Strength function}

One investigates the properties of nuclei as function of the excitation energies.
The excitation strength $S(E)$ from the initial state to the final state with the real excitation energy $E$ is expressed
in terms of the response function $R(E)$ using the relation in Eq.~(\ref{eq:delta}) as follows:
\begin{eqnarray}
\renewcommand{\arraystretch}{1.9}
S(E)
&=&{\displaystyle\sum_\nu \bra\widetilde{\Phi}_0|\hat{O}^\dagger|\Phi_\nu \ket\bra\widetilde{\Phi}_\nu|\hat{O}|\Phi_0\ket\,
    \delta(E-E_\nu)}\nonumber\\
&=&-\frac{1}{\pi}\mbox{Im}\, R(E),\\
R(E)&=&
{\displaystyle\int d\vc{r}d\vc{r}'\,\widetilde{\Phi}^*_0(\vc{r})\hat{O}^\dagger {\cal G}(E;\vc{r},\vc{r}') \hat{O} \Phi_0(\vc{r}')},
\end{eqnarray}
where $\Phi_0$, $\Phi_\nu$, and $\hat{O}$ are the initial and final states with index $\nu$ and the transition operator, respectively.
In the complex scaling, the response function is expressed as 
\begin{eqnarray}
  R^\theta(E)= \int d\vc{r}d\vc{r}'\,
  \{\widetilde{\Phi}^*_0(\vc{r})\}^\theta\hat{O}^{\dagger\,\theta} {\cal G}^\theta(E;\vc{r},\vc{r}')
  \hat{O}^\theta\Phi^\theta_0(\vc{r}'),
\end{eqnarray}
for the complex-scaled initial state $\Phi_0^\theta$ and the transition operator $\hat{O}^\theta$.
Using the decomposition of the CSGF given in Eq.~(\ref{eq:CSGF}), the complex-scaled response function can be expressed in the decomposed form:
\begin{eqnarray}
R^\theta(E)&=&R_{\rm B}^\theta(E)+R_{\rm R}^\theta(E)+R_{\rm C}^\theta(E),
\\
R_{\rm B}^\theta(E)&=&
{\displaystyle \sum_{b=1}^{N_b}\frac{\bra\widetilde{\Phi}^\theta_0|\hat{O}^{\dagger\,\theta}|\Phi_b^\theta\ket\bra\widetilde{\Phi}_b^\theta|\hat{O}^\theta|\Phi_0^\theta\ket}{E-E_b+i\epsilon}},
 \\
 R_{\rm R}^\theta(E)&=&
 {\displaystyle\sum_{r=1}^{N_r^\theta}\frac{\bra\widetilde{\Phi}^\theta_0|\hat{O}^{\dagger\,\theta}|\Phi_r^\theta\ket\bra\widetilde{\Phi}_r^\theta|\hat{O}^\theta|\Phi_0^\theta\ket}{E-E_r^{\rm res}+i\Gamma_r/2}},
 \\
 R_{\rm C}^\theta(E)&=&
 {\displaystyle\sum_{c=1}^{N_c^\theta}\frac{\bra\widetilde{\Phi}^\theta_0|\hat{O}^{\dagger\,\theta}|\Phi_c^\theta\ket\bra\widetilde{\Phi}_c^\theta|\hat{O}^\theta|\Phi_0^\theta\ket}{E-E_c^\theta}}.
 \label{eq:CSRF}
\end{eqnarray}
From the above decomposition, the strength function, $\displaystyle S^\theta(E)=-\frac{1}{\pi}$Im\,$R^\theta(E)$, is also decomposed as:
\begin{eqnarray}
S^\theta(E)&=& S_{\rm B}^\theta(E)+S_{\rm R}^\theta(E)+S_{\rm C}^\theta(E).
\label{eq:strength}
\end{eqnarray}

The matrix elements of the complex-scaled operator given in Eq.~(\ref{eq:CSRF}) are independent of $\theta$.
While $R^\theta_{\rm B}$ for bound states is independent of $\theta$,
the dependence of $R^\theta_{\rm R}$ and $R^\theta_{\rm C}$ on $\theta$ originates from $N_r^\theta$, $N_c^\theta$, and $E_c^\theta$.
On the other hand, the total strength function, $S^\theta(E)$, is an observable,
which is positive definite and independent of $\theta$, i.e. $S^\theta(E)=S(E)$,
whereas the terms of $S_{\rm R}^\theta(E)$ and $S_{\rm C}^\theta(E)$ may contain negative numbers.
The present formulation is applicable to many-body systems, in which 
the continuum term, $S_{\rm C}^\theta(E)$, is further decomposed into the different continuum states on the different $2\theta$ lines,
such as in Fig.~\ref{fig:borromean}.
From this decomposition, one can investigate the individual contributions to the strength function \cite{myo98,myo14a,myo20}.
This is a key feature of the complex scaling.

\section{Application to the two-body system}
\subsection{Schematic potential model}
\begin{figure}[b]
\begin{center}
\includegraphics[width=5.8cm,bb=0 0 396 378]{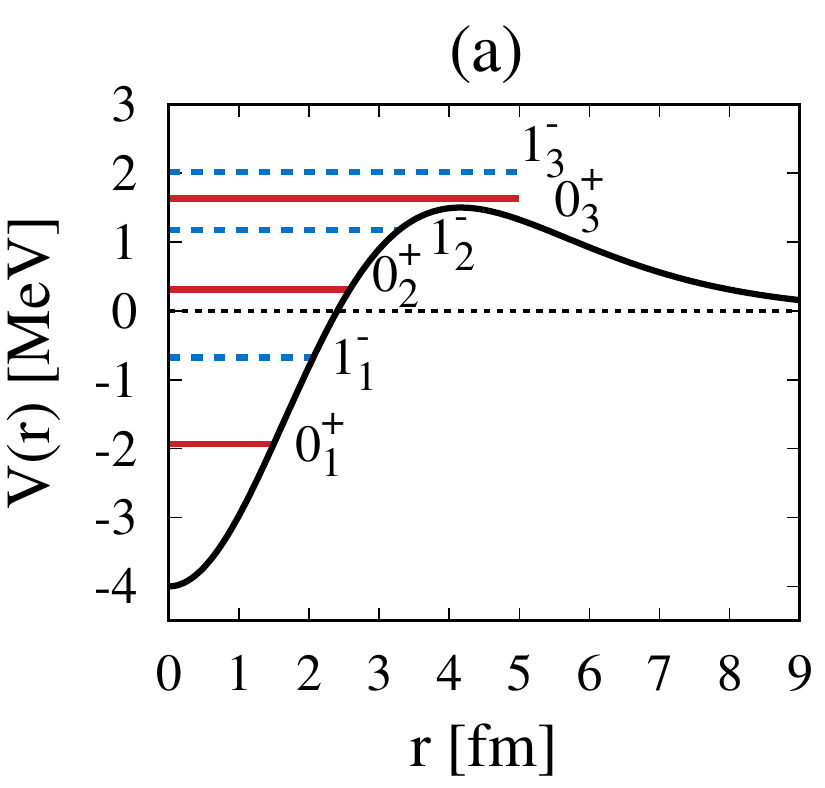}\\[0.5cm]
\includegraphics[width=8.2cm,bb=0 0 377 286]{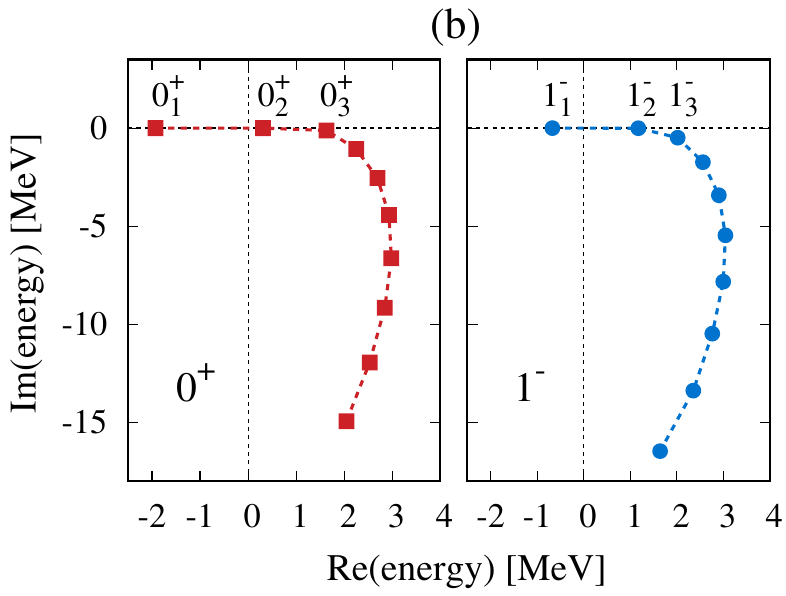}
\caption{
  (a) Potential shape and low-lying energy levels for $0^+$ and $1^-$ states. 
  (b) Discrete energy eigenvalues with complex scaling in the complex energy plane.}
\label{fig:csoto}
\end{center}
\end{figure}

We demonstrate the instructive calculations of resonances using complex scaling.
To this end, we use the schematic potential model with the simple Hamiltonian given by :
\begin{eqnarray}
H&=&-\frac{\hbar^2}{2m}\nabla^2+V(r),
\nonumber\\
V(r)&=&-8e^{-0.16r^2}+4e^{-0.04r^2}~\mbox{[MeV]},
\label{eq:csoto_pot}
\end{eqnarray}
where we set $\hbar^2/m=1$ [MeV\,fm$^2$]  \cite{homma97,myo98,csoto90,myo97}.
We calculate the spin-parity of $J^\pi=0^+$ and 1$^-$ states and evaluate the dipole transition strength. 
In this calculation, we directly solve the complex-scaled Schr\"odinger equation instead of the basis expansion.
The energy spectra are shown in Figs.~\ref{fig:csoto} (a) and (b), and we take the value of $\theta$ to be close to $\pi/4$.
Panel (a) shows the several energy levels up to the third ones for each spin, alongside the potential shape,
while panel (b) shows the distributions of the discrete energy eigenvalues in the complex energy plane. 
There is one bound state in each spin state and we obtain many resonances with various decay widths.

\begin{figure}[bh]
\centering
\includegraphics[width=7.5cm,bb=0 0 360 252]{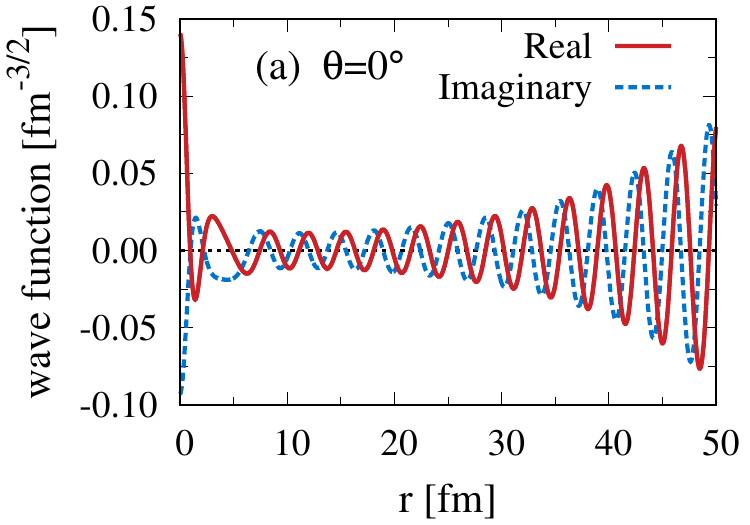}\\[0.4cm]
\includegraphics[width=7.5cm,bb=0 0 360 252]{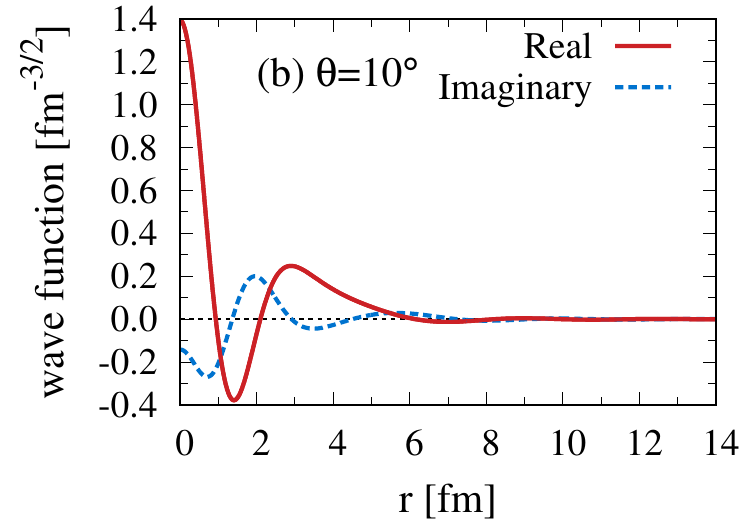}
\caption{
  Wave functions of the $0^+_3$ resonance ($E_r=1.63$ MeV, $\Gamma=0.246$ MeV) in the schematic potential model:
  (a) without and (b) with the complex scaling.}
\label{fig:WF}
\end{figure}

Figure \ref{fig:WF} (a) shows the wave function of the $0^+_3$ resonance without the complex scaling.
Its resonance energy is located at the top of the potential barrier with $E_r=1.63$ MeV and $\Gamma=0.246$ MeV, as shown in Fig.~\ref{fig:csoto}~(a).
The resonance wave function is obtained directly under the outgoing wave in the asymptotic region. 
This wave function is complex and its divergent behavior with spatial oscillation in the form of $e^{ik_{\rm R}r}$ is clearly confirmed,
although the state is not a direct observable.
Figure \ref{fig:WF} (b) shows the wave function obtained using the complex scaling with $\theta = 10^\circ$.
The damping behavior of the wave function is clearly confirmed over 8 fm.

\begin{figure}[t] 
\centering
\includegraphics[width=8.2cm,bb=0 0 450 241]{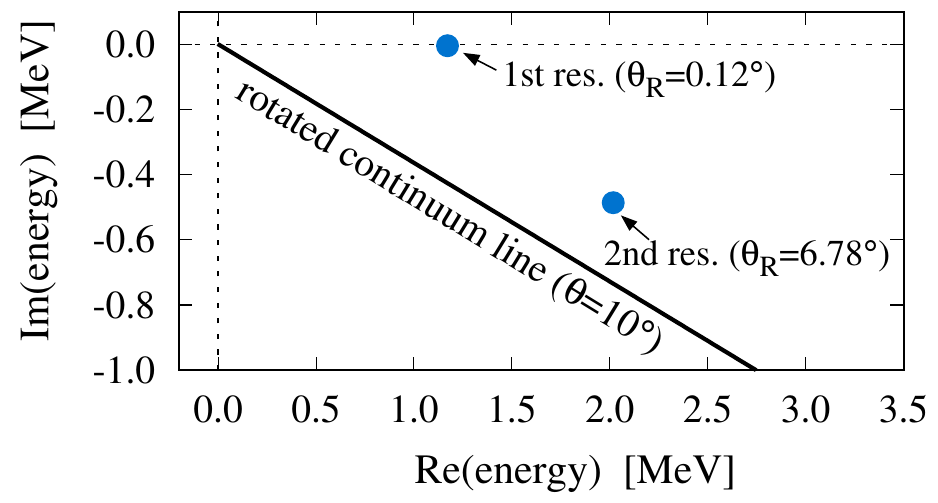}  
\caption{Energy eigenvalues of the first and second $1^-$ resonances in the complex energy plane with blue circles together with the rotated continuum line at $\theta=10^\circ$.}
\label{fig:csoto-pole}
\end{figure}

In the complex scaling, we obtain the Green's function not only of resonances but also of non-resonant continuum states.
This Green's function is used to calculate the strength functions $S(E)$, and also to solve the few-body scattering problems
\cite{suzuki05,kruppa07,odsuren21}.
We show the dipole strength function from the $0^+$ ground state to the scattering states of $1^-$ in the complex scaling,
where the transition operator is defined as $\widehat{O}=rY_{10}(\hat{r})\sqrt{4\pi}$ \cite{myo98}.
We discuss the effects of resonances and non-resonant continuum states of the $1^-$ states on the dipole strength.
As can be seen in Fig.~\ref{fig:csoto-pole}, when $\theta=10^\circ$,
we obtain one bound state ($E_0=-0.68$ MeV), the first resonance ($E_1=1.17-i0.49 \times 10^{-2}$ MeV),
and the second resonance ($E_2=2.02-i0.49$ MeV).
The residual non-resonant continuum states are on the $2\theta$ line.

\begin{figure}[t]
\centering
\includegraphics[width=8.2cm,bb=0 0 360 252]{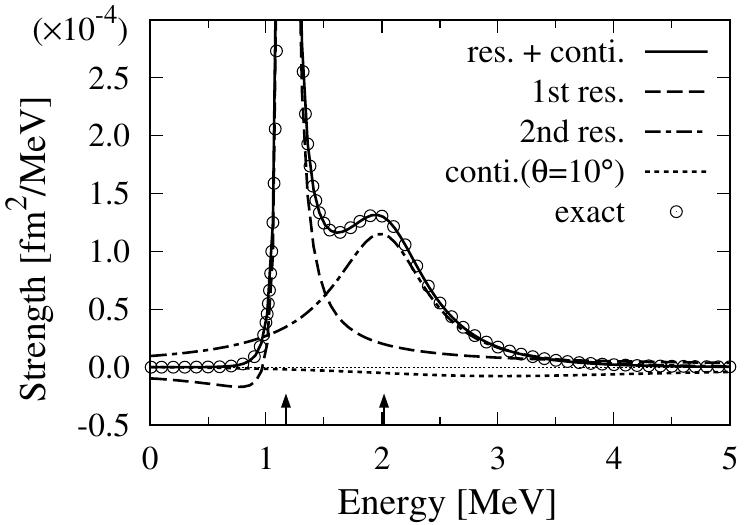} 
\caption{
  Dipole transition strength in the schematic potential model from the $0^+$ ground state to the $1^-$ unbound states 
  measured from the threshold energy with $\theta=10^\circ$ in the complex scaling.
  The dashed and dash-dotted lines are the contributions of the first and second $1^-$ resonances, respectively,
  in which their resonance energies are indicated by the vertical arrows from the bottom measure, corresponding to Fig.~\ref{fig:csoto-pole}.
  The dotted line is the continuum contribution. The solid line shows the sum of all the terms. 
  The open circles represent the exact solutions.}
\label{fig:csoto-E1}
\end{figure}

Figure~\ref{fig:csoto-E1} shows the dipole strength function.
We compare the strengths obtained using the complex scaling with the exact solutions (open circles), 
which is obtained by solving the ordinary scattering problem for the $1^-$ states.
The strength distribution is calculated using the Green's function and the dipole matrix elements using the complex scaling with $\theta=10^\circ$, 
as shown by the solid line, which agrees with the exact calculation.

In the figure, a sharp peak is observed at 1.2 MeV.
The dashed line denotes the contribution from the first resonance
and then we conclude that the sharp peak originates from the first resonance. 
The contribution of the second resonance (dash-dotted line) is found to be a component of the peak at around 2 MeV, 
whereas the contribution of the residual continuum states (dotted line), i.e. the rotated continuum states shown in Fig.~\ref{fig:csoto-pole}, 
is negligible.
This decomposition of the dipole strength makes it clear that two major peaks in the strength are due to the contributions of the two $1^-$ resonances.

\subsection{$\alpha$+$n$ system}\label{sec:CLD}

In this section, we demonstrate the application of the CLD to the realistic $\alpha$+$n$ system as the unbound nucleus $^5$He 
as an example \cite{suzuki05,odsuren14}.
For the interaction between the $\alpha$ particle ($^4$He($0^+$)) and $n$ (spin $1/2$), we use the microscopic potential \cite{kanada79}, 
consisting of the central and spin-orbit terms.
This potential is constructed from an analysis of the $\alpha$-$p$ scattering experiments,
and is used to reproduce the scattering phase shifts of the $\alpha$-$n$ system for each partial wave.
This potential is used throughout to calculate the neutron-rich He isotopes and their mirror proton-rich nuclei, explained in the next section.
The CLD is defined as the variation in the level densities due to the interaction, and the scattering phase shift
is derived from CLD. 
In terms of CLD, we can discuss the relationship between resonances and the scattering phase shifts. 

\begin{figure}[b]
\begin{center}
\includegraphics[width=8.6cm,bb=0 0 468 485]{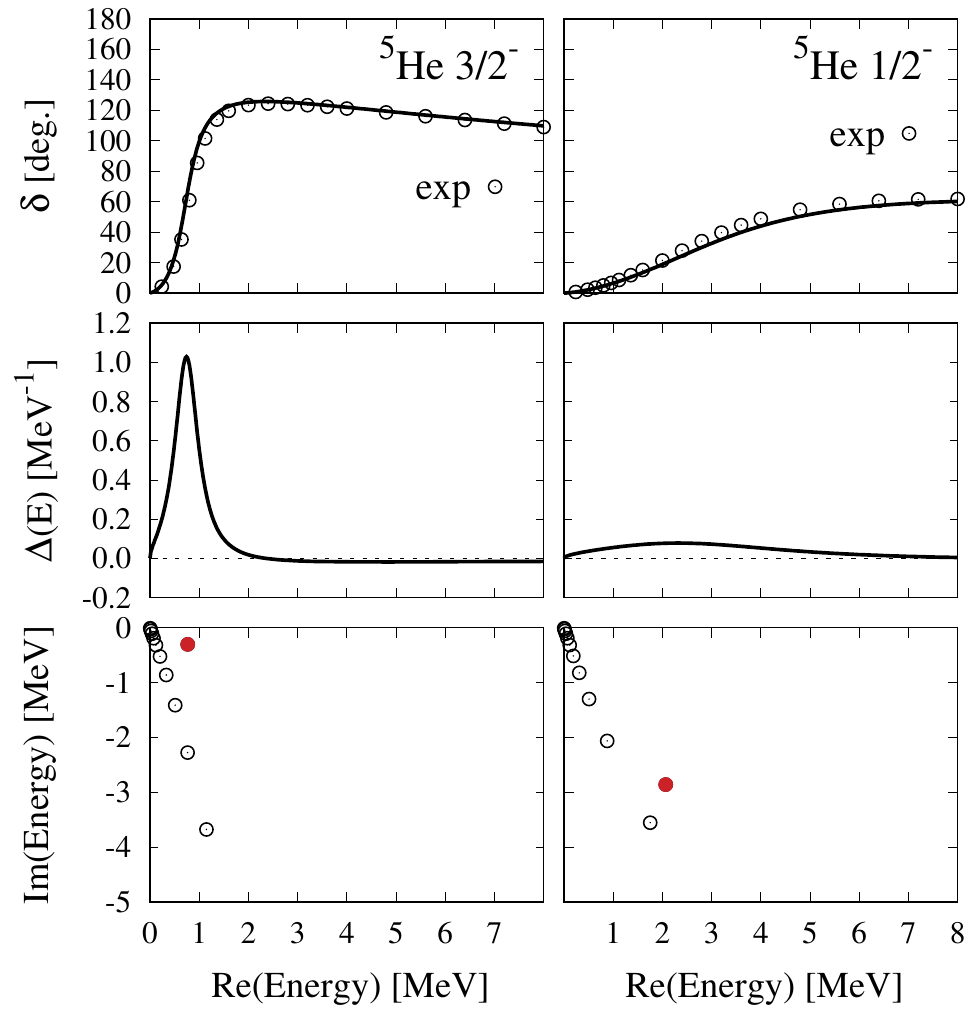}
\caption{
  Various properties of the $\alpha$+$n$ system for the $p$--wave states with $J^\pi=3/2^-$ (left) and $1/2^-$ (right). 
  Upper: phase shifts, middle: continuum level density $\Delta(E)$, and lower: energy eigenvalue distributions in a complex energy plane,
  where the resonance poles are shown by red solid circles. The scaling angle $\theta$ is 35$^\circ$.
  In the top panels, the experiments with open circles are taken from Refs. \cite{kanada79,aoyama95} and references therein.
}
\label{fig:5He_PS}
\end{center}
\end{figure}

Using the 30 Gaussian basis states, we calculate the energy eigenvalues of the complex-scaled Hamiltonian of $^5$He with $\theta=35^\circ$, and
the results for the two $p$--wave states of $3/2^-$ and $1/2^-$ are shown in Fig.~\ref{fig:5He_PS}. 
We find that the two states have one resonance pole, which corresponds to the observed resonances of $^5$He \cite{nndc}.
The resonance parameters are $(E_r,\Gamma)=(0.74,~0.59)$ MeV for the $3/2^-$ state,
and $(2.10,~5.82)$ MeV for the $1/2^-$ state.
These values in good agreement with the experimental data \cite{aoyama95,aoyama06,tilley02}.
The energy difference between the two states arises from the spin-orbit interaction between $\alpha$ and $n$.

In addition to the resonances, the discretized non-resonant continuum solutions are obtained along the $2\theta$ line as shown in Fig.~\ref{fig:5He_PS}.
We use these solutions to calculate the CLD, $\Delta(E)$, and finally obtain the phase shifts of the $\alpha$+$n$ scattering.
For $3/2^-$, $\Delta(E)$ exhibits a clear peak at the corresponding resonance energy,
and the phase shift exhibits a maximum derivative in this energy region.
For $1/2^-$, which is a broad resonance with a large decay width, the enhancement of $\Delta(E)$ is very mild and then 
the phase shift changes slowly. We can confirm the correspondence among the three quantities for each resonance.

It is noted that the CLD and phase shifts are obtained independently of the scaling angle $\theta$ in the calculation.
This property has been discussed in detail in Ref. \cite{suzuki05}.
Additionally, the results of three-body CLD with complex scaling for the $3\alpha$ model of $^{12}$C have been presented in Refs. \cite{myo14a,kurokawa24}.

\section{Many-body resonant and continuum states} \label{sec:many}

We apply the complex scaling to the phenomena of many-body resonances in unstable nuclei. 
For this purpose, we focus on the structures of neutron-rich He isotopes and their mirror proton-rich nuclei,
most states of which are unbound due to the weak-binding property of valence nucleons with respect to the $\alpha$ particle ($^4$He).
In our model, we treat the $\alpha$ particle as an inert core nucleus
and solve the motions of valence nucleons surrounding the $\alpha$ particle.

\subsection{Nuclear model} \label{sec:COSM}
\begin{figure}[b]
\centering
\includegraphics[width=7.0cm,bb=0 0 338 365]{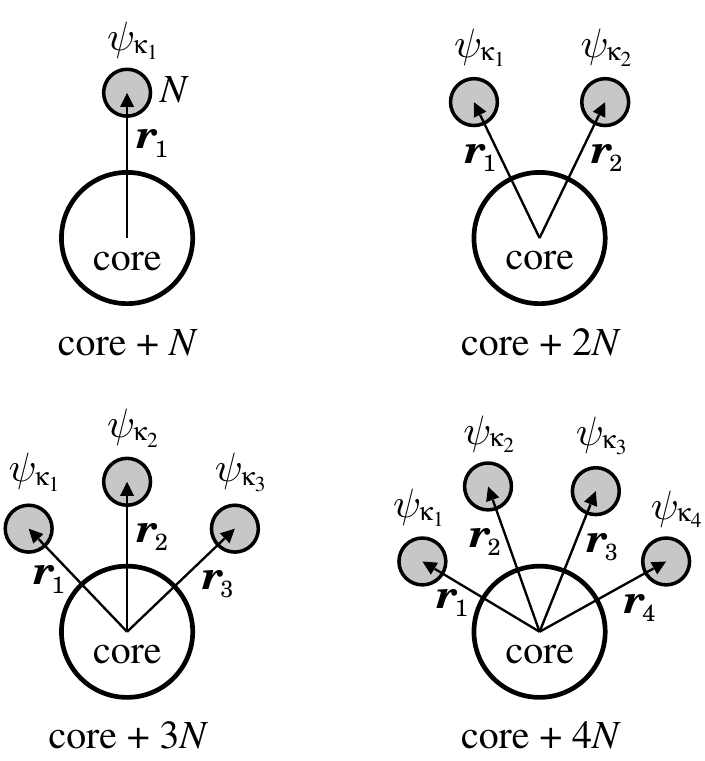}
\caption{Set of spatial coordinates consisting of a core nucleus and valence nucleons $N=n$ or $p$.}
\label{fig:COSM}
\end{figure}

We introduce a nuclear model that describes the He isotopes and their mirror nuclei \cite{suzuki88,masui06}.
Figure \ref{fig:COSM} displays the coordinate system used to solve the motions of valence nucleons around the $\alpha$ particle.
The Hamiltonian consists of the $\alpha$ particle with a mass number $A_{\rm c}=4$ and $N_{\rm v}$ valence nucleons 
with a total mass number $A=A_{\rm c}+N_{\rm v}$ \cite{masui06,myo07a,myo10}, and is given as follows:
\begin{eqnarray}
	H
&=&	\sum_{i=1}^{N_{\rm v}} t_i + t_\alpha - T_{\rm c.m.} + \sum_{i=1}^{N_{\rm v}} V^{\alpha N}_i + \sum_{i<j}^{N_{\rm v}}   V^{NN}_{ij},
        \label{eq:COSM_ham}
\end{eqnarray}
where $t_i$, $t_\alpha$, and $T_{\rm c.m.}$ are the kinetic energies of each nucleon, $\alpha$ particle, and
the center of mass of the total system, respectively.
The potentials $V^{\alpha N}$ and $V^{NN}$ are the interactions between an $\alpha$ particle and a nucleon,
and between valence nucleons, respectively, with the nuclear and Coulomb parts, where the symbol $N$ represents a neutron or a proton.

The total wave function $\Psi^J$ for a spin $J$ with a mass number $A$ is represented by a superposition of the different configurations $\Phi^{J}_{A,c}$ as follows:
\begin{eqnarray}
    \Psi^{J}
= \sum_c C^{J}_{c}\, \Phi^{J}_{c},
    \qquad
    \Phi^{J}_{c}
= \prod_{i=1}^{N_{\rm v}} a^\dagger_{\kappa_i}|0\rangle , 
    \label{eq:COSM-WF}
\end{eqnarray}
where the vacuum $|0\rangle$ represents the $\alpha$ particle with an $s$-wave $(0s)^4$ configuration of a harmonic oscillator wave function.
The creation operator $a^\dagger_{\kappa}$ is for the single-particle state of a valence nucleon above the $\alpha$ particle.
The quantum number $\kappa$ is a set of $\{n,\ell,j \}$;
the index $n$ represents the different radial component and $\ell$ is the orbital angular momentum,
which is coupled with a nucleon spin to be $j=\ell\otimes 1/2$. 
The index $c$ represents the set of $\kappa_i$ as $c=\{\kappa_1,\ldots,\kappa_{N_{\rm v}}\}$,
which determines the configuration of the valence nucleons in the many-body basis wave function $\Phi^{J}_{c}$.
The expansion coefficients $\{C_{c}^{J}\}$ in Eq.~(\ref{eq:COSM-WF}) are determined 
by solving the eigenvalue problem of the Hamiltonian matrix with the complex scaling given in Eq. (\ref{eq:eigen}).

The coordinate representation of the single-particle state $a^\dagger_{\kappa}$ is given as 
$\psi_{\kappa}(\vc{r})$ as a function of the relative coordinate $\vc{r}$ between the $\alpha$ particle 
and the valence nucleon \cite{suzuki88}.
The coordinates are explicitly shown in Fig.~\ref{fig:COSM} for up to $N_{\rm v}=4$.
We typically expand the radial part of $\psi_\kappa(\vc{r})$ with the Gaussian functions \cite{aoyama06,hiyama03} as
\begin{eqnarray}
    \psi_\kappa(\vc{r})
&=& \sum_{k=1}^{N_{\ell j}} d^k_{\kappa}\ \phi_{\ell j}^k(\vc{r}),
    \label{eq:COSM-base1}
    \\
    \phi_{\ell j }^k(\vc{r})
&=& {\cal N}\, r^{\ell} e^{-(r/b_{\ell j,k})^2/2}\, [Y_{\ell}(\hat{\vc{r}}),\chi^\sigma_{1/2}]_{j},
    \label{eq:Gauss}
	\\
    \langle \wtil \psi_\kappa|\psi_{\kappa'} \rangle 
&=& \delta_{\kappa,\kappa'},
    \label{eq:COSM-base2}
\end{eqnarray}
where we omit the isospin wave function for simplicity.
The index $k$ is to distinguish the range parameter $b_{\ell j,k}$ of the Gaussian functions with the basis number $N_{\ell j}$,
which is determined to converge the solutions.
The normalization factors of the basis functions are given by ${\cal N}$.
The coefficients $\{d^k_{\kappa}\}$ in Eq.~(\ref{eq:COSM-base1}) are determined to make
the orthonormal basis states $\{\psi_\kappa\}$ in Eq.~(\ref{eq:COSM-base2}).
It is noted that, due to the Pauli principle, the single-particle state $\psi_\kappa(\vc{r})$ in the relative $s$-wave
is imposed to be orthogonal to the $0s$ wave function in the $\alpha$ particle.
The detailed treatment is explained in Refs. \cite{aoyama01,aoyama95}.

We apply the complex scaling to the many-body wave function as
\begin{eqnarray}
    H^\theta    \, \Psi^{J,\, \theta}
&=& E_{J}^\theta\, \Psi^{J,\, \theta},
	\\
        \Psi^{J,\,\theta}
&=& \sum_c C^{J,\, \theta}_{c} \, \Phi^{J}_{c}.
\end{eqnarray}
This framework can be used for the resonance calculation of hypernuclei,
such as the neutron-rich $^9_\Lambda$He by adding a $\Lambda$ hyperon to $^8$He
to gain knowledge of the hyperon-nucleon interaction \cite{myo23d}.

\subsection{He isotopes and their mirror nuclei} \label{sec:He-COSM}

We discuss the spectroscopy of He isotopes and their mirror nuclei.
In the Hamiltonian, two kinds of the interactions between $\alpha$--$N$ and $N$--$N$ are given in Eq.~(\ref{eq:COSM_ham}).
In the present study, the nuclear interaction of $V^{\alpha N}$ is given by the microscopic potential \cite{aoyama95,kanada79},
which is used for the $\alpha$+$n$ system in the previous section.
For $V^{NN}$, we use the effective nucleon-nucleon potential \cite{tang78}, which reproduces the $s$-wave phase shifts.
We also consider the Coulomb interaction for protons.
Here we adopt the configurations $\Phi^{J}_{c}$ within $\ell \le 2$, and
slightly adjust the strength of $V^{NN}$ to reproduce the binding energy of $^6$He ($0^+$) as 0.975 MeV
measured from the $\alpha$+$n$+$n$ threshold \cite{myo21}.
This small energy helps to explain the neutron-halo structure in the ground state of $^6$He.
It is noted that the $N$-$N$ system remains unbound in this modification.

\begin{figure}[t]
\centering
\includegraphics[width=8.5cm,bb=0 0 543 343]{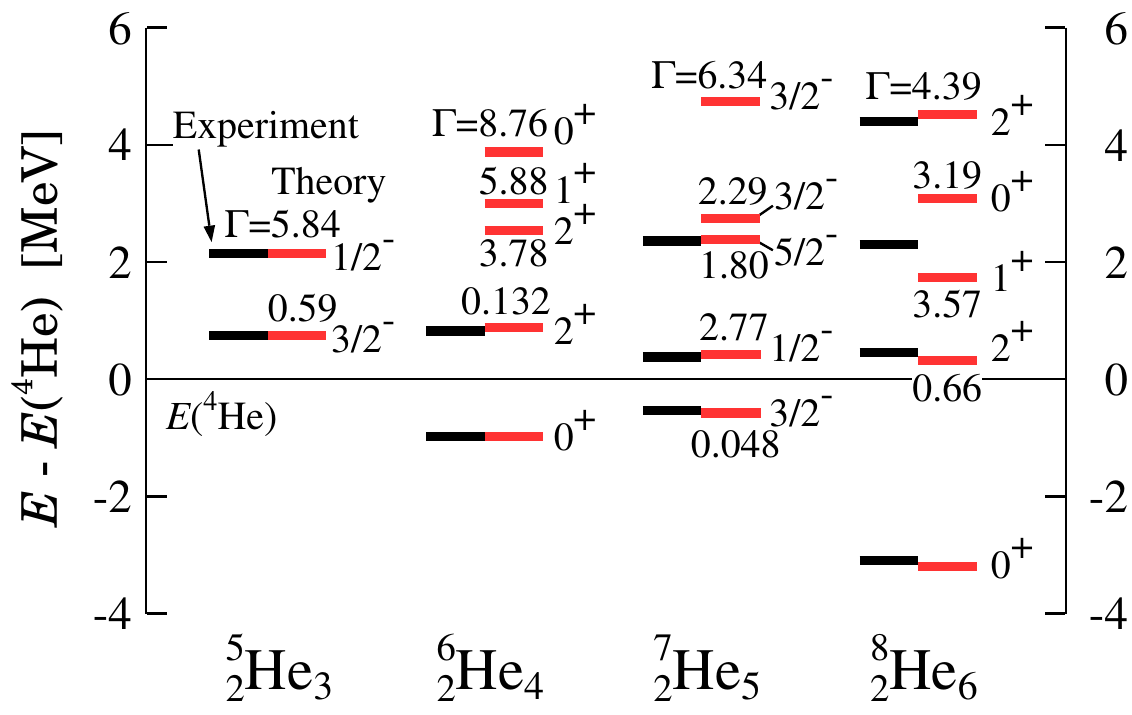}\\[0.6cm]
\includegraphics[width=8.5cm,bb=0 0 546 352]{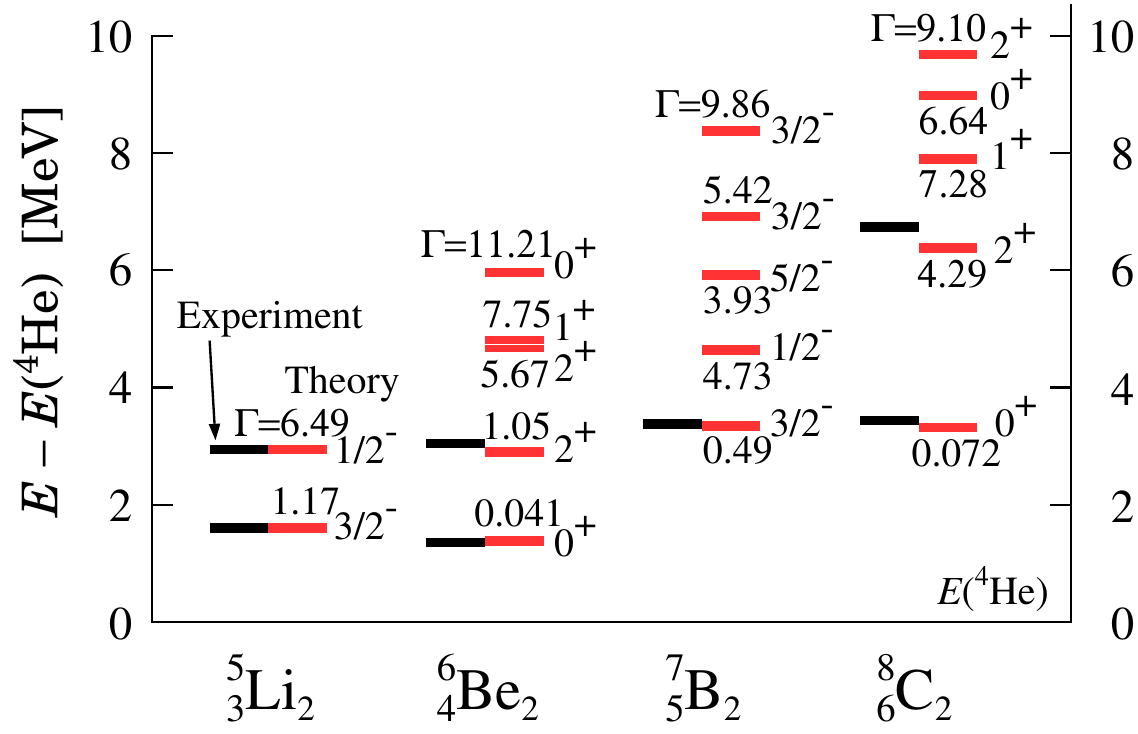}
\caption{
  Energy levels of the He isotopes (top) and their mirror nuclei (bottom) measured from the threshold energy of the $\alpha$ particle emission
  in units of MeV \cite{myo21}.
  The black (red) lines indicate the experimental (theoretical) results. The numbers near the theoretical levels represent the decay widths in MeV.}
\label{fig:ene_COSM}
\end{figure}

Figure~\ref{fig:ene_COSM} shows the energy levels of He isotopes and their mirror nuclei in 
measured from the energy of the $\alpha$ particle \cite{myo11b,myo12b,myo21}.
The numbers near the theoretical levels represent the decay widths obtained for the resonances.
There are only two bound states of $^6$He ($0^+_1$) and $^8$He ($0^+_1$). 
There is good agreement for the energy position between the experiment and theory \cite{holl21,koyama24}.
Theory also makes many predictions about the resonances of these nuclei. 
The comparison of the neutron-rich and proton-rich nuclei is interesting to understand
the mirror symmetry in nuclei and the role of the Coulomb interaction in proton-rich nuclei \cite{sobotka24,myo11b,myo14b,myo21,myo23a}.

\begin{figure}[bht]
\centering
\includegraphics[width=8.0cm,bb=0 0 360 252]{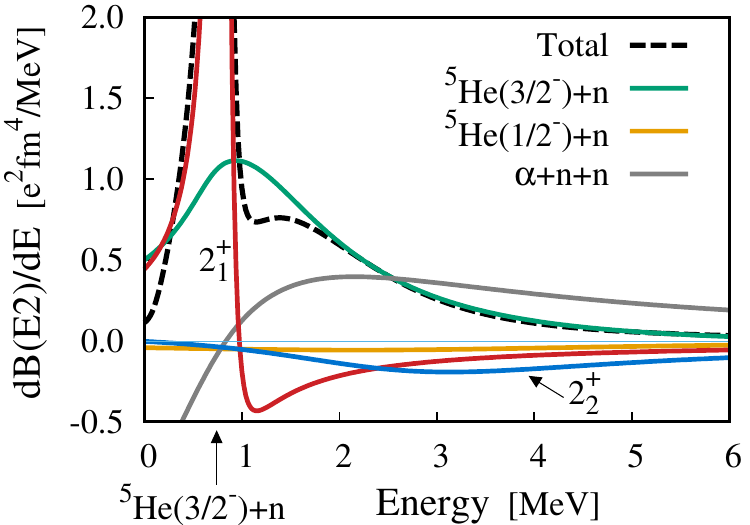}
\caption{
  Electric quadrupole ($E2$) transition of $^6$He from the ground state to the $2^+$ unbound states in the complex scaling,
  measured from the $\alpha$+$n$+$n$ threshold energy.
  The $2^+$ eigenenergies are shown in the lower panel of Fig.~\ref{fig:borromean} corresponding to the five components.
  The vertical arrow from the bottom direction indicates the threshold energy of $^5$He($3/2^-$)+$n$ channel.
}
\label{fig:6He_E2}
\end{figure}

For resonance and non-resonant continuum states in the complex scaling.
the energy eigenvalue distribution of $^6$He ($2^+$) is shown in Fig.~\ref{fig:borromean}.
In this figure, we can identify five kinds of unbound states of $^6$He : two-kinds of the three-body resonances of $2^+_1$ and $2^+_2$,
and three kinds of the continuum states of $^5$He($3/2^-$,$1/2^-$)+$n$ and $\alpha$+$n$+$n$.
Corresponding to these eigenstates, we calculate the electric quadrupole ($E2$) transition strength of $^6$He
from the $0^+$ ground state to the $2^+$ unbound states of $\alpha$+$n$+$n$ \cite{myo01}.
It is interesting to investigate the contributions of the above five components in the strength.
The $E2$ transition strength distribution is shown in Fig.~\ref{fig:6He_E2} measured from the $\alpha$+$n$+$n$ threshold energy.
As can be seen from Fig.~\ref{fig:6He_E2}, we can confirm that the $2^+_1$ resonance makes the main contribution showing
a sharp peak around the resonance energy of 0.81 MeV with a small decay width of 0.13 MeV. 
On the other hand, the contribution from the $2^+_2$ resonance is negligible due to its large decay width of 3.78 MeV.
The components associated with two- and three-body continua are smaller than that of the $2^+_1$ resonance.
However, in the continuum transitions, the two-body continuum component of $^5$He($3/2^-$)+$n$ shows 
a peak at around 1 MeV, just above the two-body threshold energy of this channel, as indicated by the vertical arrow. 
This component mainly contributes to the shoulder structure in the total strength around 1.5 MeV .
A detailed analysis of the transition strength in relation to the structure of $^6$He is provided in Ref. \cite{myo01}.

In the calculation, it is found that the $E2$ transition strength distribution 
exhibits a clear resonance behavior for the $2^+_1$ state, but not for the $2^+_2$ state.
This suggests the difficulty of the experimental observation of the $2^+_2$ state via the $E2$ transition. 
On the other hand, the continuum strengths consisting of $^5$He(3/2$^-$)+$n$ and $\alpha$+$n$+$n$ channels contribute to the formation of
a shoulder-like structure. It is also found that the total strength remains a positive definite.

\subsection{Soft dipole resonances in $^8$He and $^8$C} \label{sec:SDR}

As is mentioned in the Sect. \ref{sec:intro}, the examining the existence of the soft dipole resonances (SDR) is an interesting problem
in the field of unstable nuclear physics.
In this study, we investigate the possibility of SDR in a neutron-rich $^8_2$He$_6$ nucleus and a mirror proton-rich $^8_6$C$_2$ nucleus.
We calculate the $1^-$ states of these two nuclei and evaluate the electric dipole ($E1$) transition strength of $^8$He
from the $0^+$ bound state to the $1^-$ unbound states,
which can decay into the five-body $\alpha$+$n$+$n$+$n$+$n$ system.
The five-body ECR for $^8$He is expressed as:
\begin{eqnarray}
   1
   &=&\sum_{\nu}|\Psi^{J,\theta}_{\nu} \rangle \langle \wtil{\Psi}^{J,\theta}_{\nu}| \nonumber \\
   &=& \{\mbox{bound state of $^8$He}\}  \nonumber\\
   &+& \{\mbox{resonances of $^8$He}\}  \nonumber\\
   &+& \{\mbox{two-body continua of $^7$He$^{(*)}+n$}\} \nonumber \\
   &+& \{\mbox{three-body continua of $^6$He$^{(*)}+n+n$}\} \nonumber \\
   &+& \{\mbox{four-body continua of $^5$He$^{(*)}+n+n+n$}\} \nonumber\\
   &+& \{\mbox{five-body continua of $\alpha+n+n+n+n$}\},
   \label{eq:8He}
\end{eqnarray} 
with the state index $\nu$ \cite{myo10,myo22a,myo22b}.
This five-body ECR is useful for clarifying which states contribute to the dipole transition strength.

\begin{figure}[t]
\centering
\includegraphics[width=8.0cm,bb=0 0 360 252]{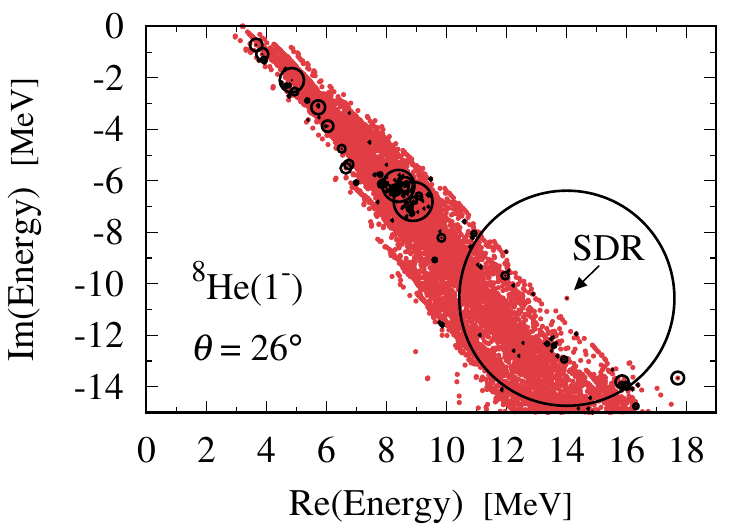}
\caption{
  Complex energies of the $1^-$ eigenstates of $^8_2$He$_6$ using the complex scaling with $\theta=26^\circ$ measured from the ground state energy.
  The magnitudes of the dipole transition matrix elements are proportional to the sizes of black circles in each state \cite{myo22a}.
  The short arrow indicates the soft dipole resonance (SDR) with $(E_x,\Gamma)=(14.0, 21.1)$ MeV having the largest dipole matrix element.
}
\label{fig:8He}
\end{figure}

Figure~\ref{fig:8He} shows the distribution of the energy eigenvalues of $^8$He ($1^-$) on the complex energy plane
as a function of the excitation energy ($E_x$), measured from the ground state of $^8$He \cite{myo22a,myo22b}.
We set $\theta = 26^{\circ}$ and the discretized continuum states are obtained almost along the $2\theta$ lines 
starting from the energy eigenvalues of the subsystems such as of $\alpha+n+n+n+n$ with 3.2 MeV. 
A resonance pole is confirmed at $(E_x,~\Gamma) = (14.0,~21.1)$ MeV.
In the figure, the size of the black open circle is proportional to the magnitude of the matrix element of the dipole transition
for that eigenstate of $^8$He.
The dipole resonance accounts for approximately half of the transition strength from the ground state \cite{myo22a,myo22b}.  
Additionally, a detailed analysis of the transition density reveals that
the four neutrons ($4n$) are spatially confined around the $\alpha$ particle, explained in Ref.~\cite{myo22a}.
This property is consistent with the physical picture of the dipole oscillation of $4n$ with respect to the $\alpha$ particle
as shown in Fig.~\ref{fig:soft}.
Therefore we regard this resonance as the soft dipole resonance.

\begin{figure}[th]
\centering
\includegraphics[width=8.0cm,bb=0 0 360 252]{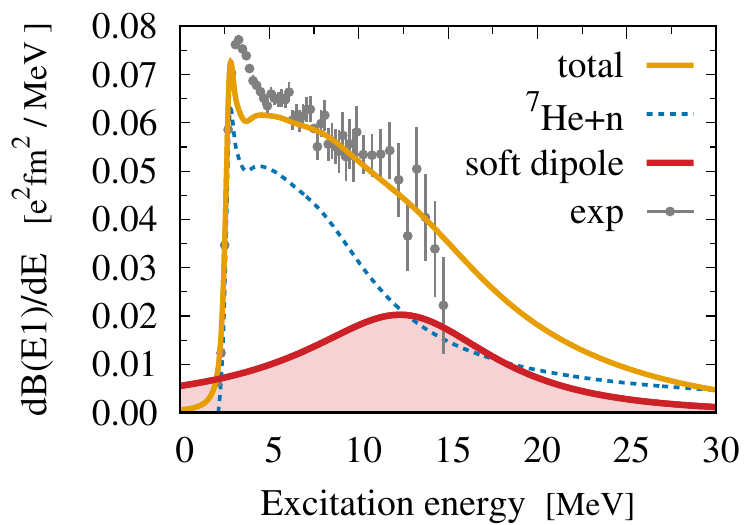} 
\caption{
  Electric dipole strength of $^8$He using the complex scaling as a function of the excitation energy $E$ with a yellow solid line,
  taken from Ref. \cite{myo22b}.
  The decomposed components are
  $^7$He(3/2$^-$)+$n$ with a blue dashed line and the soft dipole resonance with a red solid line.
  The experimental data is taken from Ref. \cite{lehr21} multiplying the factor of 0.7.
}
\label{fig:cmp}
\end{figure}

Figure~\ref{fig:cmp} shows the dipole strength function of $^8$He obtained using the complex-scaled Green's function
using Eq.~(\ref{eq:strength}) \cite{myo22a},
as a function of the excitation energy on the real energy axis.
In the total strength (yellow solid line), the sharp peak at 2.5 MeV comes from the $^7$He+$n$ two-body continuum component
just above the corresponding threshold energy,
while the soft dipole resonance creates a broad bump structure around 14 MeV in the distribution due to its large decay width of 21 MeV.
We compare the total strength with the experimental data extracted from the Coulomb breakup experiment of $^8$He \cite{lehr21}.
The energy dependence of the theory and experiment are in agreement with each other.

\begin{figure}
\centering
\includegraphics[width=8.0cm,bb=0 0 360 252]{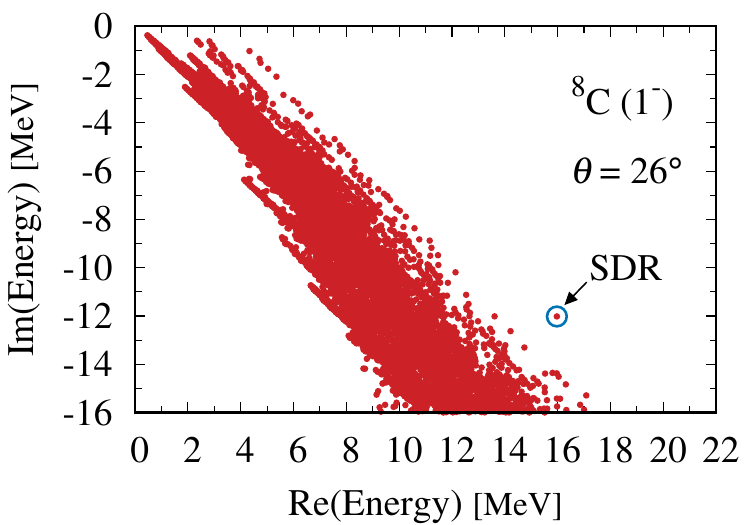} 
\caption{
   $1^-$ eigenstates of $^8_6$C$_2$ obtained by using the complex scaling with $\theta =26^{\circ}$
  in the complex energy plane measured from the $\alpha+p+p+p+p$ threshold energy.
  The blue circle indicates the eigenenergy of the soft dipole resonance (SDR) with $(E_x,\Gamma)=(12.7,~24.0)$ MeV \cite{myo23b}.}
\label{fig:8C}       
\end{figure}

We also examine the possibility of SDR in $^8_6$C$_2$, the mirror nucleus of $^8_2$He$_6$ by exchanging protons and neutrons.
In this case, $^8$C is an unbound nucleus that can decay into the $\alpha$+$p$+$p$+$p$+$p$ five-body system.
Figure~\ref{fig:8C} shows the energy eigenvalues of the $1^-$ states of $^8$C measured from the threshold energy of $\alpha+p+p+p+p$
using the same scaling angle $\theta=26^\circ$ as used for $^8$He. 
We confirm the resonance pole at $(E_r,~\Gamma) = (16.0,~24.0)$ MeV, which corresponds to a mirror state of the SDR of $^8$He. 

It is interesting to compare the spatial properties of the two SDRs of $^8$He and $^8$C as well as their ground states,
in order to discuss the mirror symmetry in neutron-rich and proton-rich nuclei.
When the $A=8$ system is divided into the $\alpha$ particle and $4N$ ($4n$ or $4p$),
the root-mean-square radius of the total system consists of the $\alpha$ particle part,
the $4N$ part, and the relative distance between them.
Here, the radius of $4N$ is measured from the center-of-mass position of $4N$ ($\vc{r}_{4N}$) within the $A=8$ system
and is defined as $\sqrt{\langle \sum_{i=1}^4 (\vc{r}_i-\vc{r}_{4N})^2 \rangle/4}$,
as shown in Fig.~\ref{fig:COSM}.

Table \ref{tab:radius} lists these values for four states of $^8$He and $^8$C.
It is noted that, for resonances, the radius can be a complex number,
and we recently proposed a new scheme to interpret the complex expectation values for resonances~\cite{myo23b},
in which the real part is related to the physical meaning, as is discussed in the Sect. \ref{sec:expect}.
According to this scheme, we discuss the real part of the complex radius.

In the table, only the ground state of $^8$He is a bound state and other three states are resonances.
Using real parts, the SDRs demonstrate larger radii than those of the ground states in the two nuclei.
For $^8$He, the radius of $4n$ in the SDR is 3.72 fm, which is larger than 2.91 fm in the ground state. 
The distance between the $\alpha$ particle and $4n$ also increases from 2.05 fm in the ground state to 2.67 fm in the SDR.
These results suggest that in the SDR, the size of the $4n$ state increases and 
the relative distance is extended because of the dipole oscillation of $4n$ with respect to the $\alpha$ particle.

For the mirror symmetry in the ground states of $^8$He and $^8$C,
the radius of $^8$C (2.81 fm) is larger than that of $^8$He (2.53 fm) due to the Coulomb repulsion of valence protons in $^8$C.
The $\alpha$--$4p$ distance (2.36 fm) in $^8$C is also larger than the $\alpha$--$4n$ distance (2.05 fm) in $^8$He. 
The mean radius of $4p$ (3.29 fm) in $^8$C is about 0.4 fm larger than that of $4n$ (2.91 fm) in $^8$He.
These differences indicate the breaking of the symmetry in their ground states due to the Coulomb interaction in $^8$C. 
For SDR, it is found that their radii (3.11 fm and 3.09 fm) show similar values in the two nuclei and
this tendency could also be confirmed in the other components of $4N$ and $\alpha$--$4N$.
These results indicate a good symmetry in the SDRs of $^8$He and $^8$C, which is different to that observed in their ground states.
A detailed discussion can be found in Ref. \cite{myo23a}.

Our prediction of the SDRs in $^8$He and $^8$C would be examined in future experiments.
The present results also suggest general existence of the SDR in both the neutron-rich and proton-rich nuclei,
which would be worthy for the understanding of nuclear mirror symmetry.

\begin{table}
\centering
\caption{Spatial sizes of the $0^+$ ground states (gr) and the $1^-$ soft dipole resonances (SDR) 
  in the mirror nuclei $^8_2$He$_6$ and $^8_6$C$_2$,
  with the $\alpha$ particle plus four valence nucleons ($4N$), where $N=n$ or $p$ \cite{myo22b}. Units are in fm.
  The values of radius, $4N$, and $\alpha$--$4N$ represent the radii of the total system and the $4N$ part,
  and the relative distance between $\alpha$ and $4N$, respectively. 
}
\label{tab:radius}
\renewcommand{\arraystretch}{1.50}
\begin{tabular}{rccccccc}
\hline
	            && \multicolumn{2}{c}{$^8_2$He$_6$} &\;& \multicolumn{2}{c}{$^8_6$C$_2$} \\ \cline{3-4} \cline{6-7}
	            && $0^+_{\rm gr}$ & $1^-_{\rm SDR}$ && $0^+_{\rm gr}$ & $1^-_{\rm SDR}$ \\ \hline	 
radius (fm)         && $2.53$~        & $3.11+0.86i$    && $2.81-0.08i$~  & $3.09+1.27i$ \\
$4N$ (fm)           && $2.91$~        & $3.72+1.14i$    && $3.29-0.13i$~  & $3.71+1.70i$ \\
$\alpha$--$4N$ (fm) && $2.05$~        & $2.67+0.84i$    && $2.36-0.03i$~  & $2.62+1.15i$ \\ \hline
\end{tabular}
\end{table}

\section{Complex expectation values associated with resonances} \label{sec:expect}

It is known in general that the expectation values of a Hermitian operator for resonances can be complex numbers,
the physical interpretation of which is a long-standing problem
\cite{berggren96,burgers96,homma97,sekihara13,dote18,myo14b,michel22}.
There is an attempt to evaluate the variance of the expectation value of an operator using its imaginary part \cite{berggren96}. 
However this discussion is limited to the operators that can commute with the Hamiltonian.
A recent idea is that the expectation values for resonances are supposed to be the function of the complex energy eigenvalue of $E_r-i\Gamma/2$.
If $\Gamma \ll E_r$, the expectation values can be expanded in a Taylor series at $E_r$ with the imaginary part being proportional to $\Gamma$
\cite{michel22}.

In this review, we present our recent idea on a possible scheme to interpret the complex expectation values for resonances \cite{myo23b}.
We utilize the Green's function of resonances as used for the strength function.
We formulate a general expression for the use of complex expectation values in the strength function, and
discuss the roles of the real and imaginary parts of the expectation values in the structure of the strength function.
This formulation is similar to the Morimatsu-Yazaki method \cite{morimatsu94}, 
which is used to calculate the energy spectrum of the hadron formation in the two-body scattering.

We apply the present scheme to the calculation of radii of resonances,
the interpretation of which has been discussed in various fields of physics \cite{burgers96,homma97,sekihara13,dote18,myo14b}.

\subsection{Formulation}
We consider the two-body system with a single channel and define the resonance wave function $\Phi_{\rm R}$
as the decaying state \cite{gamow28,siegert39}, which has a complex eigenenergy  of $E_r-i\Gamma/2$.
We then express the extended completeness relation (ECR) \cite{berggren68} 
using the solutions of the bound (B), resonant (R), and non-resonant continuum ($E_c$) states.
\begin{eqnarray}
  1      &=& \sum_{\rm B} \kets{\Phi_{\rm B}}\bras{\wtil \Phi_{\rm B}} + \sum_{\rm R} \kets{\Phi_{\rm R}}\bras{\wtil \Phi_{\rm R}}
          +  \int d E_c \kets{\Phi_{E_c}}\bras{\wtil \Phi_{E_c}}
          \nonumber\\
          &=& \sum_{~\nu}\hspace{-0.5cm}\int~\kets{\Phi_\nu}\bras{\wtil \Phi_\nu},
\end{eqnarray}
where $\nu$ is the unified index for both the discrete and continuous states.
As in the transition case, we introduce the expectation value $M_{\rm EV}$ of an arbitrary Hermitian operator $\hat O$ for the resonance,
which is complex and defined as follows:
\begin{eqnarray}
  M_{\rm EV}&=& \bra \wtil \Phi_{\rm R} | \hat O|\Phi_{\rm R} \ket ~=~ M_{\rm R} +i M_{\rm I} .
\label{eq:ev}
\end{eqnarray}
where $M_I\neq 0$.
We express the strength function $S(E)$ for the expectation value using the Green's function as
\begin{eqnarray}
	S(E) 
&=&	\sum_{~\nu}\hspace{-0.5cm}\int~
	\bras{\wtil{\Psi}_\nu}\hO\kets{\Psi_\nu}\,
	\delta(E-E_\nu)
=       \sum_{~\nu}\hspace{-0.5cm}\int~ S_\nu(E) ,
        \label{eq:strength5}
        \\
        S_\nu(E) 
&=&     -\frac1{\pi}\ {\rm Im}\left\{  \frac{
	\bras{\wtil{\Psi}_\nu} \hO  \kets{\Psi_\nu  }
        }{E^+-E_\nu}
        \right\} . 
        \label{eq:strength6}
\end{eqnarray}
The resonance contribution $S_{\rm R}(E)$ is written as 
\begin{eqnarray}
	S_{\rm R}(E)
        &=&     -\frac1{\pi}\ {\rm Im}\left\{  \frac{M_{\rm EV} }{E-E_r + i\Gamma/2} \right\}
        \\
        &=&  \frac1{\pi} \frac{M_{\rm R}\Gamma/2 - M_{\rm I}(E-E_r)}{(E-E_r)^2+\Gamma^2/4} .
        \label{eq:strength7}
\end{eqnarray}
The integration of $S_{\rm R}(E)$ over the energy gives the real part of the expectation value for resonance.
\begin{eqnarray}
        \int_{-\infty}^\infty S_{\rm R}(E)\  dE
        &=& M_{\rm R} .
\end{eqnarray}
The roles of $M_{\rm R}$ and $M_{\rm I}$ in the strength function $S_{\rm R}(E)$ in Eq.~(\ref{eq:strength7})
can be summarized as follows:
\begin{enumerate}

\item
  The real part, $M_{\rm R}$, determines the amount of the expectation value for resonance,
  which corresponds to the integration of the strength function.
  In the term including $M_{\rm R}$ in Eq.~(\ref{eq:strength7}),
  the strength distribution follows the Breit-Wigner form with the centroid energy $E_r$.

\item
  The imaginary part, $M_{\rm I}$, produces a deviation from the Breit-Wigner distribution, which is an odd function measured from $E_r$.
  The energy at the peak of the strength function $S_{\rm R}(E)$ shifts from $E_r$ due to $M_{\rm I}$.

\item
  The energy $E_{\rm max}$ at the maximum strength and the energy $E_{\rm min}$ at the minimum strength are given, respectively, as:
\begin{eqnarray}
E_{\rm max} &=&  E_r + \frac{\Gamma}{2} \frac{ M_{\rm R} - |M_{\rm EV}| }{M_{\rm I}} ,
\label{eq:max_E}
\\
E_{\rm min} &=&  E_r + \frac{\Gamma}{2} \frac{ M_{\rm R} + |M_{\rm EV}| }{M_{\rm I}} .
\label{eq:min_E}
\end{eqnarray}
The strength function becomes zero at the energy of $E_r + \Gamma/2 \cdot M_{\rm R}/M_{\rm I}=(E_{\rm max}+E_{\rm min})/2$,
which is a middle point between $E_{\rm max}$ and $E_{\rm min}$.
\end{enumerate}

\begin{figure}[b]
\centering
\includegraphics[width=8.0cm,bb=0 0 360 252]{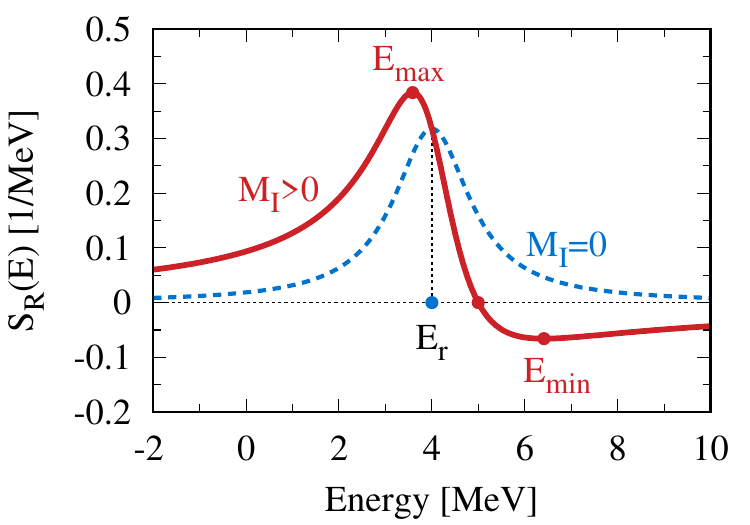}
\caption{
  Schematic illustration of the arbitrary strength distribution for resonance, $S_{\rm R}(E)$,
  as a function of the excitation energy $E$ shown by the red solid line,
  where $E_r=4$ MeV, $\Gamma=2$ MeV, $M_{\rm R}=M_{\rm I}=1$ (arb. unit).
  The blue dotted line shows the Breit-Wigner distribution with an imaginary part of zero ($M_{\rm I}=0$).
  }
\label{fig:strength}
\end{figure}

Figure~\ref{fig:strength} shows the typical behavior of the strength function $S_{\rm R}(E)$.
From these formulas, the role of the imaginary part $M_{\rm I}$ in $M_{\rm EV}$ can be understood in the strength function.
In actual calculations, the total strength function $S(E)$ is not only affected by resonances, but also by the non-resonant continuum states.
It should be noted that the total strength $S(E)$ is observable and can be positive definite depending on the operators such as radius operator.
On the other hand, the component $S_\nu(E)$ is not necessarily positive definite,
because the resonance components are not observable and are permitted to exhibit negative values at certain energies \cite{myo98,myo14a}.
This property can also be seen in the transition strength, as is demonstrated in Sect.~\ref{sec:many}.

\subsection{Results of $^{12}$C and $^6$He} \label{sec:result}

We present the strength distribution of the square radius ($r^2$) of resonances,
a topic that has been explored in both molecular physics \cite{burgers96} and hadron physics \cite{sekihara13,dote18}.
We select two nuclei, $^{12}$C with the $3\alpha$ cluster model and $^6$He with the $\alpha$+$n$+$n$ model,
and examine the strength functions of $r^2$ for resonances in the   two nuclei.

\begin{table}[t]
  \caption{
    Resonance energies $E_r$, decay widths $\Gamma$, root-mean-square radii $\sqrt{\EV{r^2}}$ of the resonances of $^{12}$C ($3\alpha$) and $^6$He,
    taken from Refs. \cite{kurokawa07,myo23b},
    The values are measured from the threshold energies of $\alpha+\alpha+\alpha$ and $\alpha+n+n$, respectively.
    We also show two bound states with real negative energies for reference.
  }
\label{tab:ene}
\centering
\renewcommand{\arraystretch}{1.50}
\begin{tabular}{cc|cccc}
            &  State   & $E_r$ (MeV)     & $\Gamma$ (MeV)    & $\sqrt{\EV{r^2}}$ (fm)  \\ \hline
 $^{12}$C   &  $0^+_1$ & $-7.29$         &  --               &  $2.36$ \\  
            &  $0^+_2$ & $0.76$          & $2.4\times 10^{-3}$  &  $4.23+0.49i$ \\  
            &  $0^+_3$ & $1.66$          & $1.48$  &  -- \\  
            &  $0^+_4$ & $4.58$          & $1.1$   &  $3.49+0.75i$ \\  
            &  $0^+_5$ & $14.3$          & $1.5$   &  $2.89+0.20i$ \\  
 \hline
 $^6$He     &  $0^+_1$ & $-0.975$        & --      & $2.37$  \\  
            &  $2^+_1$ & $0.879$         & $0.132$ &  $3.05+1.39i$  \\  
            &  $2^+_2$ & $2.52$          & $3.78$  &  $4.54+3.59i$ \\  
 \hline
\end{tabular}
\end{table}

Table \ref{tab:ene} lists the resonance energies $E_r$, the decay widths $\Gamma$, and the radii for the resonances in the two nuclei
taken from \cite{kurokawa07,myo23b}. The complex eigenenergies are obtained using the complex scaling. 
For reference, we also list two bound states of $^{12}$C, $^6$He with negative real energies. 
In $^{12}$C, the $0^+_3$ state is a resonance, the energy of which is obtained
using the analytical continuation method combined with complex scaling for extrapolation.
Therefore, the radius of this resonance is not reported in Ref. \cite{kurokawa07}.

\begin{figure*}[t]
\centering
\includegraphics[width=18.0cm,bb=0 0 1012 254]{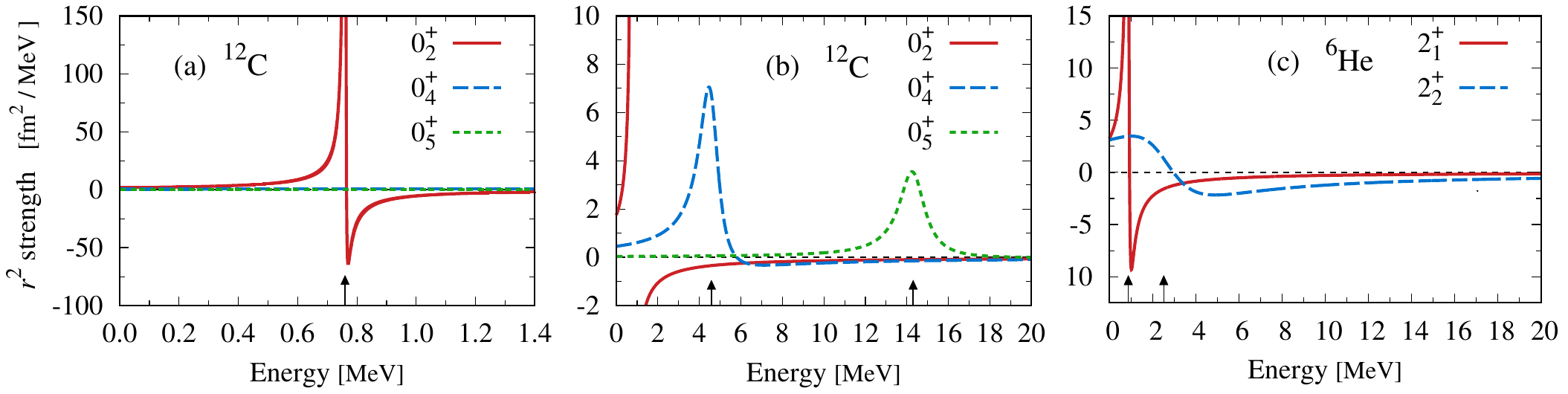} 
\caption{
  Strength functions of the square radii of $^{12}$C and $^6$He, taken from Ref. \cite{myo23b}.
  (a) and (b):~$^{12}$C, $0^+_2$, $0^+_4$, and $0^+_5$ measured from the $\alpha+\alpha+\alpha$ threshold energy.
  (c):~$^6$He, $2^+_1$ and $2^+_2$ measured from the $\alpha+n+n$ threshold energy.
  Units of the vertical axis are in fm$^2$/MeV. The short vertical arrows from the bottom measures indicate the resonance energies $E_r$
  listed in Table \ref{tab:ene}.
}\label{fig:12C}
\end{figure*}

Figures~\ref{fig:12C} (a) and \ref{fig:12C} (b) show the $r^2$ strength functions
for the $0^+_2$, $0^+_4$, and $0^+_5$ resonances of $^{12}$C
measured from the $\alpha+\alpha+\alpha$ threshold energy, with different energy scales.
For the $0^+_2$ resonance, known as a famous Hoyle state, having a gas like structure of a $3\alpha$ system \cite{horiuchi12},
and the $r^2$ strength shows a very sharp peak at the resonance energy due to the very small decay width of 2.4 keV \cite{kurokawa07}.
Due to the imaginary part of $\sqrt{\EV{r^2}}$, a negative component appears just above the resonance energy of 0.76 MeV.
This negative region will be covered by contributions from the non-resonant continuum states showing the positive-valued distribution.
For $0^+_4$, the distribution shows a peak at an energy slightly lower than the resonance energy shown by the arrow, by 0.12 MeV.
One can confirm the deviation of the shape from the Breit-Wigner type.
For $0^+_5$, the distribution shows the small peak with an energy close to the resonance energy.
The shape of the distribution resembles a Breit-Wigner type due to the small imaginary part of $\sqrt{\EV{r^2}}$ for this state,
as shown in Table \ref{tab:ene}.

Figure \ref{fig:12C} (c) shows the $r^2$ strength functions of $^{6}$He with
two resonances of $2^+_1$ and $2^+_2$, measured from the threshold energy of $\alpha+n+n$.
In the two resonances, $2^+_1$ has a small decay width of 0.132 MeV and this state is well confirmed by the experiments \cite{nndc,tilley02}.
For the $2^+_1$ state, the $r^2$ strength exhibits a sharp peak at the resonance energy with a large deviation from the Breit-Wigner shape,
due to the imaginary part of $\sqrt{\EV{r^2}}$ of this resonance as detailed in Table \ref{tab:ene}.
For the $2^+_2$ state showing large decay widths, the distributions also largely deviate from the Breit-Wigner shape.

We confirm a variety of the distributions in the $r^2$ strength functions for resonances in two nuclei
due to the presence of the imaginary part of the $r^2$ expectation values.
The present scheme can be the basis to evaluate the effects of resonances on physical quantities in the real-energy distribution.

\section{Summary and perspective}

We have explained some of the interesting aspects of resonance phenomena in light unstable nuclei.
Most of them become many-body resonances corresponding to the many-body threshold due to the weak binding of valence nucleons
with respect to a core nucleus.
We introduced a powerful method of the complex scaling to describe many-body resonances and continuum states.
In this review, we presented the applications of the complex scaling to nuclear many-body systems
with a focus on the light neutron-rich and proton-rich nuclei.
This method can be applicable to the other quantum many-body systems as well as the atomic nuclei.

The fundamental concept of the complex scaling is that resonances can be expressed as the square-integrable $L^2$ functions.
Based on this property, the bound, resonant, and continuum states can be described simultaneously by solving the eigenvalue problem
of the complex-scaled Hamiltonian.
This method can also be used to describe many-body unbound states straightforwardly.
The bi-orthogonal solutions obtained, such as resonances and anti-resonances,
form a complete set, and the Green's function is then constructed in terms of these eigensolutions.
The Green's function provides a kind of projection of the matrix elements of the complex energy states
onto the physical quantities observed at real energies.
These quantities can be decomposed into the contributions from each resonance as well as from different kinds of continuum states.
This concept can be used to interpret the complex expectation values associated with resonances.

The complex scaling can provide a unified description of the nuclear structure and reactions, although there are still unsolved problems.
The complex scaling describes resonances with complex energy eigenvalues, the imaginary part of which represents the total decay width.
It is important to extract the partial decay widths of many-body resonances for each decaying channel,
which provide useful information on the decay properties of resonances, but they are not yet available.

In addition to the resonances, the virtual $s$-wave states are a kind of the unbound state and often plays an important role
in nuclear structure and reactions around threshold energies \cite{ito25},
but the virtual states are still difficult to obtain directly in the complex scaling with $\theta<\pi/2$, which is different from the resonance poles.
For example, the neutron halo structure of $^{11}$Li originates from a large mixing of the $s$-wave component of the valence neutrons.
This property is related to the presence of the $s$-wave virtual states in the $^9$Li+$n$ system \cite{masui00}, a subsystem of $^{11}$Li.
However, it is unclear how to identify the three-body virtual states as members of the three-body eigenstates of $^{11}$Li. 
A similar problem is seen in $^9$Be in the $\alpha$+$\alpha$+$n$ three-body system.
The importance of the $s$-wave virtual state in the $^8$Be+$n$ channel has been indicated, in which $^8$Be is a resonance
\cite{efros99,odsuren15,kikuchi16,odsuren17}.
This state can affect the dipole transition strength from the ground state,
which is important for understanding the production rate of $^9$Be in the universe from an astrophysical point of view.
It would be beneficial to develop a theoretical framework that can treat many-body virtual states in many-body unbound systems.

Interpreting the complex expectation values for resonances is a long-standing problem. 
We have presented a scheme that provides a possible interpretation of the complex numbers in terms of the Green's function \cite{myo23b}.
This clarifies the roles of the real and imaginary parts in the expectation values.
It is, however, still difficult to verify this scheme because the resonance poles and the relevant matrix elements are not directly observables.
Further development and consideration of this problem would be necessary.

Recently, the observation of multineutron systems such as $3n$ and $4n$ has been developed \cite{marques02,kisamori16,duer22}.
Theoretically, the possibility of the existence of multineutron resonances remains a topic of discussion 
\cite{pieper03,lazauskas05,shirokov16,hiyama16,gandolfi17,fossez17,deltuva19,li19,ishikawa20,higgins20,lazauskas23,wan25},
with various approaches including the complex scaling, but a definitive conclusion has yet to be reached.

\section*{Acknowledgements}
The author would like to thank Prof. Kiyoshi Kat\=o for his valuable comments that helped to complete the manuscript.

\section*{Authors' contributions}
TM wrote the draft and completed the writing of the manuscript.
He proofread and approved the final version of the manuscript.

\section*{Funding}
The work shown in this paper was supported by JSPS KAKENHI Grants No. JP22K03643, No. JP25H01268, No. JP26K07092, and JST ERATO Grant No. JPMJER2304, Japan.
Numerical calculations were partly achieved through the use of SQUID at the Cybermedia Center, Osaka University.

\section*{Data availability}
The data that support the findings of this study are available from the corresponding author upon reasonable request.

\section*{Declarations}
\subsection*{Competing interests}
The authors declare that they have no competing interests.

\bibliographystyle{apsrev4-1} 
\bibliography{reference_AAPPS2} 

\end{document}